\documentclass[longauth]{aa}  

\usepackage[utf8]{inputenc}
\usepackage{graphicx}
\usepackage{txfonts}
\usepackage{euclid}
\usepackage{hyperref}
\usepackage[dvipsnames]{xcolor} 
\usepackage[normalem]{ulem}

\begin{document}

   \title{\Euclid preparation: X. The \Euclid photometric-redshift challenge}

\author{Euclid Collaboration: G.~Desprez$^{1}$\thanks{\email{Guillaume.Desprez@unige.ch}}, S.~Paltani$^{1}$, J.~Coupon$^{1}$, I.~Almosallam$^{2,3,4}$, A.~Alvarez-Ayllon$^{1}$, V.~Amaro$^{5}$, M.~Brescia$^{6}$, M.~Brodwin$^{7}$, S.~Cavuoti$^{6,8,9}$, J.~De Vicente-Albendea$^{10}$
% ...
}

\institute{$^{1}$ Department of Astronomy, University of Geneva, ch. d'\'Ecogia 16, CH-1290 Versoix, Switzerland\\
$^{2}$ Saudi Information Technology Company, Riyadh 12382, Saudi Arabia\\
$^{3}$ King Abdulaziz City for Science and Technology, Riyadh 11442, Saudi Arabia\\
$^{4}$ Information Engineering, Parks Road, Oxford OX1 3PJ, UK\\
$^{5}$ School of Physics and Astronomy, Sun Yat-sen University, Guangzhou 519082, Zhuhai Campus, China\\
$^{6}$ INAF-Osservatorio Astronomico di Capodimonte, Via Moiariello 16, I-80131 Napoli, Italy\\
$^{7}$ Department of Physics and Astronomy, University of Missouri, 5110 Rockhill Road, Kansas City, MO 64110, USA\\
$^{8}$ Department of Physics "E. Pancini", University Federico II, Via Cinthia 6, I-80126, Napoli, Italy\\
$^{9}$ INFN section of Naples, Via Cinthia 6, I-80126, Napoli, Italy\\
$^{10}$ Centro de Investigaciones Energ\'eticas, Medioambientales y Tecnol\'ogicas (CIEMAT), Avenida Complutense 40, 28040 Madrid, Spain\\
$^{11}$ School of Physics, HH Wills Physics Laboratory, University of Bristol, Tyndall Avenue, Bristol, BS8 1TL, UK\\
$^{12}$ Department of Physics, Oxford University, Keble Road, Oxford OX1 3RH, UK\\
$^{13}$ Aix-Marseille Univ, CNRS, CNES, LAM, Marseille, France\\
$^{14}$ McWilliams Center for Cosmology, Department of Physics, Carnegie Mellon University, Pittsburgh, PA 15213, USA\\
$^{15}$ Center for Astrophysics | Harvard \& Smithsonian, 60 Garden St., Cambridge, MA 02138, USA\\
$^{16}$ INAF-Osservatorio Astronomico di Roma, Via Frascati 33, I-00078 Monteporzio Catone, Italy\\
$^{17}$ Universit\"ats-Sternwarte M\"unchen, Fakult\"at f\"ur Physik, Ludwig-Maximilians-Universit\"at M\"unchen, Scheinerstrasse 1, 81679 M\"unchen, Germany\\
$^{18}$ Sorbonne Universit{\'e}s, UPMC Univ Paris 6 et CNRS, UMR 7095, Institut d'Astrophysique de Paris, 98 bis bd Arago, 75014 Paris, France\\
$^{19}$ Max Planck Institute for Extraterrestrial Physics, Giessenbachstr. 1, D-85748 Garching, Germany\\
$^{20}$ INAF-Osservatorio Astronomico di Brera, Via Brera 28, I-20122 Milano, Italy\\
$^{21}$ INAF-Osservatorio di Astrofisica e Scienza dello Spazio di Bologna, Via Piero Gobetti 93/3, I-40129 Bologna, Italy\\
$^{22}$ SISSA, International School for Advanced Studies, Via Bonomea 265, I-34136 Trieste TS, Italy\\
$^{23}$ INFN, Sezione di Trieste, Via Valerio 2, I-34127 Trieste TS, Italy\\
$^{24}$ INAF-Osservatorio Astronomico di Trieste, Via G. B. Tiepolo 11, I-34131 Trieste, Italy\\
$^{25}$ Universidad de la Laguna, E-38206, San Crist\'{o}bal de La Laguna, Tenerife, Spain\\
$^{26}$ Instituto de Astrof\'{i}sica de Canarias. Calle V\'{i}a L\`{a}ctea s/n, 38204, San Crist\'{o}bal de la Laguna, Tenerife, Spain\\
$^{27}$ Dipartimento di Fisica e Astronomia, Universit\'a di Bologna, Via Gobetti 93/2, I-40129 Bologna, Italy\\
$^{28}$ INFN-Sezione di Bologna, Viale Berti Pichat 6/2, I-40127 Bologna, Italy\\
$^{29}$ IFPU, Institute for Fundamental Physics of the Universe, via Beirut 2, 34151 Trieste, Italy\\
$^{30}$ INAF-Osservatorio Astrofisico di Torino, Via Osservatorio 20, I-10025 Pino Torinese (TO), Italy\\
$^{31}$ INFN-Sezione di Roma Tre, Via della Vasca Navale 84, I-00146, Roma, Italy\\
$^{32}$ Department of Mathematics and Physics, Roma Tre University, Via della Vasca Navale 84, I-00146 Rome, Italy\\
$^{33}$ Instituto de Astrof\'isica e Ci\^encias do Espa\c{c}o, Universidade do Porto, CAUP, Rua das Estrelas, PT4150-762 Porto, Portugal\\
$^{34}$ Dipartimento di Fisica e Scienze della Terra, Universit\'a degli Studi di Ferrara, Via Giuseppe Saragat 1, I-44122 Ferrara, Italy\\
$^{35}$ INAF, Istituto di Radioastronomia, Via Piero Gobetti 101, I-40129 Bologna, Italy\\
$^{36}$ Institut de Recherche en Astrophysique et Plan\'etologie (IRAP), Universit\'e de Toulouse, CNRS, UPS, CNES, 14 Av. Edouard Belin, F-31400 Toulouse, France\\
$^{37}$ INFN-Sezione di Torino, Via P. Giuria 1, I-10125 Torino, Italy\\
$^{38}$ Dipartimento di Fisica, Universit\'a degli Studi di Torino, Via P. Giuria 1, I-10125 Torino, Italy\\
$^{39}$ Universit\'e C\^ote d'Azur, Observatoire de la C\^ote d'Azur, CNRS, Laboratoire Lagrange, Bd de l'Observatoire, CS 34229, 06304 Nice cedex 4, France\\
$^{40}$ INAF-IASF Milano, Via Alfonso Corti 12, I-20133 Milano, Italy\\
$^{41}$ Institut de F\'{i}sica d’Altes Energies (IFAE), The Barcelona Institute of Science and Technology, Campus UAB, 08193 Bellaterra (Barcelona), Spain\\
$^{42}$ Instituto de Astrof\'isica e Ci\^encias do Espa\c{c}o, Faculdade de Ci\^encias, Universidade de Lisboa, Tapada da Ajuda, PT-1349-018 Lisboa, Portugal\\
$^{43}$ Institute of Space Sciences (ICE, CSIC), Campus UAB, Carrer de Can Magrans, s/n, 08193 Barcelona, Spain\\
$^{44}$ Institut d’Estudis Espacials de Catalunya (IEEC), 08034 Barcelona, Spain\\
$^{45}$ AIM, CEA, CNRS, Universit\'{e} Paris-Saclay, Universit\'{e} Paris Diderot, Sorbonne Paris Cit\'{e}, F-91191 Gif-sur-Yvette, France\\
$^{46}$ Observatoire de Sauverny, Ecole Polytechnique F\'ed\'erale de Lau- sanne, CH-1290 Versoix, Switzerland\\
$^{47}$ INAF-Osservatorio Astrofisico di Arcetri, Largo E. Fermi 5, I-50125, Firenze, Italy\\
$^{48}$ Centre National d'Etudes Spatiales, Toulouse, France\\
$^{49}$ Institute for Astronomy, University of Edinburgh, Royal Observatory, Blackford Hill, Edinburgh EH9 3HJ, UK\\
$^{50}$ University of Nottingham, University Park, Nottingham NG7 2RD, UK\\
$^{51}$ European Space Agency/ESRIN, Largo Galileo Galilei 1, 00044 Frascati, Roma, Italy\\
$^{52}$ ESAC/ESA, Camino Bajo del Castillo, s/n., Urb. Villafranca del Castillo, 28692 Villanueva de la Ca\~nada, Madrid, Spain\\
$^{53}$ Univ Lyon, Univ Claude Bernard Lyon 1, CNRS/IN2P3, IP2I Lyon, UMR 5822, F-69622, Villeurbanne, France\\
$^{54}$ University of Lyon, UCB Lyon 1, CNRS/IN2P3, IUF, IP2I Lyon, France\\
$^{55}$ Departamento de F\'isica, Faculdade de Ci\^encias, Universidade de Lisboa, Edif\'icio C8, Campo Grande, PT1749-016 Lisboa, Portugal\\
$^{56}$ Instituto de Astrof\'isica e Ci\^encias do Espa\c{c}o, Faculdade de Ci\^encias, Universidade de Lisboa, Campo Grande, PT-1749-016 Lisboa, Portugal\\
$^{57}$ Universit\'e Paris-Saclay, CNRS, Institut d'astrophysique spatiale, 91405, Orsay, France\\
$^{58}$ Department of Physics \& Astronomy, University of Sussex, Brighton BN1 9QH, UK\\
$^{59}$ Astrophysics Group, Blackett Laboratory, Imperial College London, London SW7 2AZ, UK\\
$^{60}$ INFN-Bologna, Via Irnerio 46, I-40126 Bologna, Italy\\
$^{61}$ Institut de Physique Nucl\'eaire de Lyon, 4, rue Enrico Fermi, 69622, Villeurbanne cedex, France\\
$^{62}$ Aix-Marseille Univ, CNRS/IN2P3, CPPM, Marseille, France\\
$^{63}$ Department of Physics, P.O. Box 64, 00014 University of Helsinki, Finland\\
$^{64}$ Department of Physics and Helsinki Institute of Physics, Gustaf H\"allstr\"omin katu 2, 00014 University of Helsinki, Finland\\
$^{65}$ Dipartimento di Fisica "Aldo Pontremoli", Universit\'a degli Studi di Milano, Via Celoria 16, I-20133 Milano, Italy\\
$^{66}$ INFN-Sezione di Milano, Via Celoria 16, I-20133 Milano, Italy\\
$^{67}$ Mullard Space Science Laboratory, University College London, Holmbury St Mary, Dorking, Surrey RH5 6NT, UK\\
$^{68}$ Institute of Theoretical Astrophysics, University of Oslo, P.O. Box 1029 Blindern, N-0315 Oslo, Norway\\
$^{69}$ Jet Propulsion Laboratory, California Institute of Technology, 4800 Oak Grove Drive, Pasadena, CA, 91109, USA\\
$^{70}$ von Hoerner \& Sulger GmbH, Schlo{\ss}Platz 8, D-68723 Schwetzingen, Germany\\
$^{71}$ Max-Planck-Institut f\"ur Astronomie, K\"onigstuhl 17, D-69117 Heidelberg, Germany\\
$^{72}$ Institut d'Astrophysique de Paris, 98bis Boulevard Arago, F-75014, Paris, France\\
$^{73}$ Universit\'e de Gen\`eve, D\'epartement de Physique Th\'eorique and Centre for Astroparticle Physics, 24 quai Ernest-Ansermet, CH-1211 Gen\`eve 4, Switzerland\\
$^{74}$ NOVA optical infrared instrumentation group at ASTRON, Oude Hoogeveensedijk 4, 7991PD, Dwingeloo, The Netherlands\\
$^{75}$ Argelander-Institut f\"ur Astronomie, Universit\"at Bonn, Auf dem H\"ugel 71, 53121 Bonn, Germany\\
$^{76}$ Institute for Computational Cosmology, Department of Physics, Durham University, South Road, Durham, DH1 3LE, UK\\
$^{77}$ Institut f\"ur Theoretische Physik, University of Heidelberg, Philosophenweg 16, 69120 Heidelberg, Germany\\
$^{78}$ Zentrum f\"ur Astronomie, Universit\"at Heidelberg, Philosophenweg 12, D- 69120 Heidelberg, Germany\\
$^{79}$ Universit\'e C\^ote d'Azur, Observatoire de la C\^ote d’Azur, CNRS, Laboratoire Lagrange, France\\
$^{80}$ INAF-IASF Bologna, Via Piero Gobetti 101, I-40129 Bologna, Italy\\
$^{81}$ Universit\'e de Paris, F-75013, Paris, France, LERMA, Observatoire de Paris, PSL Research University, CNRS, Sorbonne Universit\'e, F-75014 Paris, France\\
$^{82}$ Space Science Data Center, Italian Space Agency, via del Politecnico snc, 00133 Roma, Italy\\
$^{83}$ Institute of Space Science, Bucharest, Ro-077125, Romania\\
$^{84}$ Institute for Computational Science, University of Zurich, Winterthurerstrasse 190, 8057 Zurich, Switzerland\\
$^{85}$ INFN-Padova, Via Marzolo 8, I-35131 Padova, Italy\\
$^{86}$ Dipartimento di Fisica e Astronomia “G.Galilei", Universit\'a di Padova, Via Marzolo 8, I-35131 Padova, Italy\\
$^{87}$ Departamento de F\'isica, FCFM, Universidad de Chile, Blanco Encalada 2008, Santiago, Chile\\
$^{88}$ Universidad Polit\'ecnica de Cartagena, Departamento de Electr\'onica y Tecnolog\'ia de Computadoras, 30202 Cartagena, Spain\\
$^{89}$ Infrared Processing and Analysis Center, California Institute of Technology, Pasadena, CA 91125, USA\\
$^{90}$ Jodrell Bank Centre for Astrophysics, School of Physics and Astronomy, University of Manchester, Oxford Road, Manchester M13 9PL, UK\\
$^{91}$ Department of Physics and Astronomy, University College London, Gower Street, London WC1E 6BT, UK\\
}

   \date{Received date; accepted date}

 
\abstract{
   Forthcoming large photometric surveys for cosmology require precise and accurate photometric redshift (photo-$z$) measurements for the success of their main science objectives. However, to date, no method has been able to produce photo-$z$s at the required accuracy using only the broad-band photometry that those surveys will provide. An assessment of the strengths and weaknesses of current methods is a crucial step in the eventual development of an approach to meet this challenge.
   We report on the performance of 13 photometric redshift code single value redshift estimates and redshift probability distributions (PDZs) on a common set of data, focusing particularly on the $0.2$--$2.6$  redshift range that the \textit{Euclid} mission will probe.
   We designed a challenge using emulated \textit{Euclid} data drawn from three photometric surveys of the COSMOS field. The data was divided into two samples: one calibration sample for which photometry and redshifts were provided to the participants; and the validation sample, containing only the photometry to ensure a blinded test of the methods. Participants were invited to provide a redshift single value estimate and a PDZ for each source in the validation sample, along with a rejection flag that indicates the sources they consider unfit for use in cosmological analyses. The performance of each method was assessed through a set of informative metrics, using cross-matched spectroscopic and highly-accurate photometric redshifts as the ground truth.
   We show that the rejection criteria set by participants are efficient in removing strong outliers, that is to say sources for which the photo-$z$ deviates by more than $0.15(1+z)$ from the spectroscopic-redshift (spec-$z$). We also show that, while all methods are able to provide reliable single value estimates, several machine-learning methods do not manage to produce useful PDZs. We find that no machine-learning method provides good results in the regions of galaxy color-space that are sparsely populated by spectroscopic-redshifts, for example $z>1$. However they generally perform better than template-fitting methods at low redshift ($z<0.7$), indicating that template-fitting methods do not use all of the information contained in the photometry.
   We introduce metrics that quantify both photo-$z$ precision and completeness of the samples (post-rejection), since both contribute to the final figure of merit of the science goals of the survey (e.g., cosmic shear from \textit{Euclid}). Template-fitting methods provide the best results in these metrics, but we show that a combination of template-fitting results and  machine-learning results with rejection criteria can outperform any individual method. On this basis, we argue that further work in identifying how to best select between machine-learning and template-fitting approaches for each individual galaxy should be pursued as a priority.
   }

    \authorrunning{Euclid Collaboration: G. Desprez et al.}
   \maketitle
%

\section{Introduction}

\label{sec:intro}

The estimation of galaxy redshifts through their photometry, or photometric redshifts (photo-$z$s), has evolved significantly since the concept was first proposed. 
The earliest attempts to determine redshifts from photometry used empirical relations (e.g., \citealt{Baum1962,Loh1986,Connolly1995}), which then evolved into template-fitting of the photometry (e.g., \citealt{Puschell1982,Koo1985,Lanzetta1996,Arnouts1999,Bolzonella2000}). More recently, machine-learning algorithms have been used, based purely on photometry (e.g., \citealt{Firth2003,Tagliaferri2003,Collister2004}), possibly combining photometric and morphological information (e.g., \citealt{Way2009,Singal2011,Gomes2018,Soo2018}), and even directly fed with image cutouts of the sources (e.g., \citealt{D'Isanto2018,Pasquet2019}). Photometric redshifts were first used to complement spectroscopic-redshifts (spec-$z$) when the latter were not available, and they subsequently became a major tool used in modern cosmological surveys to compute redshifts for large numbers of sources. 
For instance, the Dark Energy Survey (DES; \citealt{Flaugher2005}), the Kilo-Degree Survey (KiDS; \citealt{deJong2013}), the Hyper Suprime Cam Subaru Strategic Program (HSC-SSP; \citealt{Aihara2018}), the \textit{Euclid} survey \citep{Laureijs2011}, the Vera C. Rubin Observatory Legacy Survey of Space and Time (LSST; \citealt{Ivezic2019}), and the \textit{Roman Space Telescope} survey \citep{Akeson2019} all rely or will rely on photo-$z$s to carry out their main science goals. \citet{Salvato2019} present a review of the various ways to compute photo-$z$s and the challenges these large surveys face.

For cosmological applications, the quality of photo-$z$ measurements is important, since constraints on cosmological parameters obtained by photometric surveys depend on their precision and their accuracy. The performance of photo-$z$ determination depends on several factors (e.g., the set of filters and their depths, the quality of the photometry, the correction of observational effects, etc.), among which the algorithm plays a key role. For this reason, a large variety of photo-$z$ codes have been developed using different approaches for the problem, and a great deal of work is ongoing to improve these methods.

Tests comparing the results of several methods can be carried out to assess state-of-the-art algorithm performance and to identify possible improvements. Such tests have been performed on different sets of data, including: \citet{Hogg1998} on the \textit{Hubble} Deep Field North; \citet{Hildebrandt2010} on the photo-$z$ Accuracy Testing (PHAT) contest based on simulations and Great Observatories Origins Deep Survey (GOODS; \citealt{Giavalisco2004}) data; \citet{Abdalla2011} on the SDSS-DR6 Luminous Red Galaxies sample; \citet{Dahlen2013} on the Cosmic Assembly Near-infrared Deep Extragalactic Legacy Survey (CANDELS; \citealt{Grogin2011}); \citet{Tanaka2018} on the HSC-SSP data-release 1; and \citet{Schmidt2020} on simulated data.

The \textit{Euclid} survey \citep{Laureijs2011} is a large photometric and spectroscopic survey, which is planned to cover 15\,000\,deg$^{2}$ of the northern and southern extragalactic sky with a 1.2\,m-diameter space telescope. \textit{Euclid}'s main goal is to investigate the Universe's accelerating expansion through two main probes, baryonic acoustic oscillations and weak-lensing tomography. The latter probe requires determination of the shapes and redshifts of galaxies. The measurement of source shapes will be performed using a wide visible band ($VIS$) covering 540--920\,nm. For the determination of the photo-$z$s, \textit{Euclid} will also perform near-infrared (NIR) photometric observations, in $Y$, $J$, and $H$ bands (960--2000\,nm), complemented by optical ground-based external observations (EXT) in $u$, $g$, $r$, $i$ and $z$ bands. \citet{Laureijs2011} present the requirements that the \textit{Euclid} photo-$z$s must meet in order to achieve the desired figure of merit (FoM) for the science goals. The choice of the methods to derive the photo-$z$s is driven by these requirements. Therefore, the \textit{Euclid} photo-$z$ team has designed a test for photo-$z$ methods using a photometry and filter set defined specifically for \textit{Euclid}. Several photo-$z$ codes, most of them being developed by members of the \textit{Euclid} Collaboration, have been applied to a realistic set of \textit{Euclid}-like data obtained from images of the COSMOS field \citep{Scoville2007}. This field was chosen because of its large collection of spectroscopic-redshifts, required by machine-learning algorithms to perform efficiently. 

As in \citet{Hogg1998} and \citet{Hildebrandt2010}, we have set up a blind test of the performance of the photo-$z$ methods. They are evaluated using standard estimators, as well as new estimators defined specifically for \textit{Euclid}. However, because the final complementary optical-photometry data sets are expected to be deeper and cover a broader wavelength range in the late stages of the \textit{Euclid} mission than those available here, we do not expect to meet the photo-$z$ requirements; we can only compare the relative performance of the different algorithms. In this challenge, we focus on the precision (the scatter) of the results and the fraction of catastrophic failures, but not on the accuracy (the bias) of the photo-$z$s. We assume that the \textit{Euclid} photo-$z$s can be calibrated and therefore that the bias can be removed, for instance using the complete calibration of the color-redshift relation (C3R2; \citealt{Masters2015,Masters2017,Masters2019,Guglielmo2020}). The precision of photo-$z$s is nevertheless extremely important for the success of tomographic analyses, because the scatter makes the bins overlap in true-redshift space; hence the larger the scatter, the larger the degeneracy between the weak-lensing signal in the different bins. If too large, this degeneracy would effectively prevent us from studying the evolution of the dark-energy properties across the different epochs of the Universe.

The point of this challenge is first to help the \textit{Euclid} photo-$z$ team to define the strategy of the \textit{Euclid} photo-$z$ pipeline to achieve the photo-$z$ requirements. It also aims to provide clues on ways to improve photo-$z$ method performance by comparing the pros and cons of different approaches.


\section{Data}
\label{sec:data}

We built a \textit{Euclid}-like wide-survey data set from real photometric data matched as far as possible to the characteristics of the future \textit{Euclid} survey. However, some unavoidable differences are present. First the broad \textit{VIS} band does not exist in any other survey and thus cannot be simulated from existing data. Also, the ground-based optical data we have used (the DES survey; see Sect.~\ref{sec:data-img-ext}) do not contain any $u$-band observations, although we expect to have such observations over most of the \textit{Euclid} survey. In addition, deeper ground-based data are expected to be available in the late stage of the survey. Finally, the available NIR images have a lower resolution than \textit{Euclid} will have. For these reasons, this challenge cannot be interpreted as a test of the absolute performance of the photo-$z$ codes, but only as a comparison of the different algorithm under similar conditions. 

\subsection{Images}
\label{sec:data-img}

\begin{figure}
    \centering
    \includegraphics[width=\linewidth]{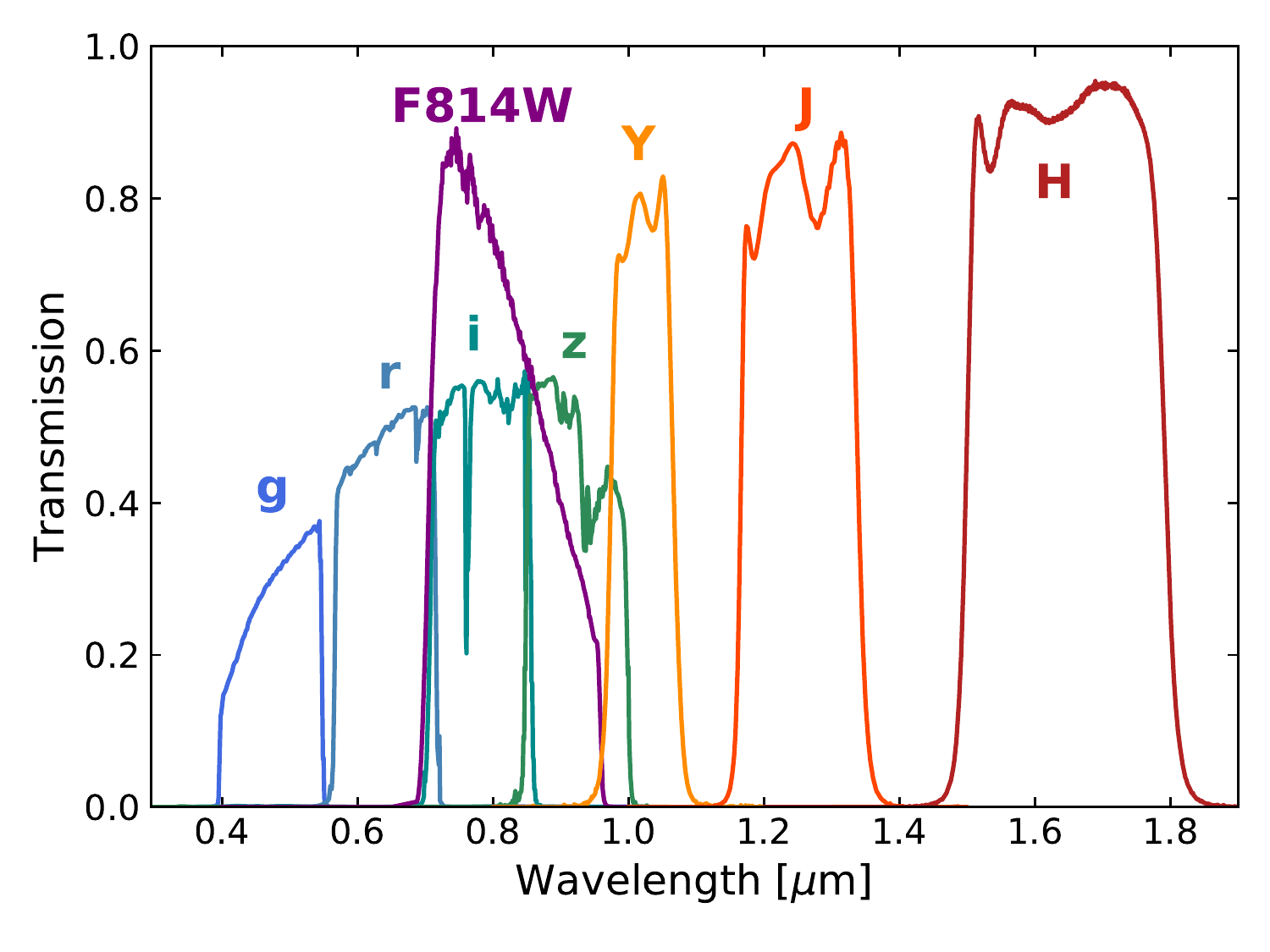}
    \caption{Transmission curves of the eight filters used in the challenge. The effects of instrumental throughput and atmosphere are included in the transmission.}
    \label{fig:transmission}
\end{figure}

The data set is composed of mosaics in eight different bands ($g$, $r$, $i$, $z$, $Y$, $J$, $H$, and $VIS$-like) from three different surveys of the COSMOS field. The area covered by the images is $\sim$1.2$\times$1.2\,deg$^{2}$. 
The transmission curves of the filters in these bands are shown in Fig.~\ref{fig:transmission}.
All the mosaics have been rescaled to the same pixel scale (i.e., $ \ang{;;0.1}$ per pixel). Table~\ref{tab:extLike} shows the properties of the different mosaics.

\begin{table*}
    \centering
    \caption{Properties of the Ext-like, NIR-like, and $VIS$-like images used to generate the ``\textit{Euclid}'' mosaics.}
    \begin{tabular}{rcccccccc}
    \hline\hline
    \rule{0pt}{1.2em}& PSF-FWHM  & Depth  & Native pixel scale   & Zero-point & Error correction factor \\
    & [arcsec] & [AB mag, 10~$\sigma$] & [arcsec pixel$^{-1}$] & [AB mag] & \\
    \hline
    \rule{0pt}{1.2em}$g$ & 1.250 & 24.20 & 0.27 & 31.90 & 1.247\\
     $r$ & 1.151 & 23.85 & 0.27 & 32.32 & 1.259 \\
     $i$ & 1.005 & 22.96 & 0.27 & 30.19 & 1.380 \\
     $z$ & 0.807 & 22.45 & 0.27 & 31.26 & 1.191 \\
     $Y$ & 0.855 & 23.81 & 0.15 & 30.00 & 2.884 \\
     $J$ & 0.831 & 23.59 & 0.15 & 30.00 & 2.582 \\
     $H$ & 0.800 & 23.13 & 0.15 & 30.00 & 2.377 \\
     $VIS$-like & 0.200 & 24.50 & 0.03 & 25.49 & 1.038\\
    \hline
    \end{tabular}
    \label{tab:extLike}
\end{table*}

\subsubsection{VIS-like image}
\label{sec:data-img-vis}

The $VIS$-like mosaic has been emulated using ACS F814W images\footnote{\url{http://irsa.ipac.caltech.edu/data/COSMOS/images/acs_mosaic_2.0/}} acquired by the \textit{Hubble Space Telescope} (\textit{HST}, \citealt{Koekemoer2007,Massey2010}).
It has been generated by re-binning and smoothing the \textit{HST} images to the \textit{Euclid} pixel scale and resolution ($ \ang{;;0.1}/$px), and adding random Gaussian noise to match the planned \textit{Euclid} $VIS$ depth. The zero-point determination is described in \citet{Bohlin2016}.

Both the scientific image and rms map have been created using dedicated simulation software (see Appendix~\ref{sec:visSoft} for more information).
Although the $VIS$-like image has the required depth and resolution, the F814W filter ($0.7$--$0.95\,\micron$) is narrower than the $VIS$ one ($0.54$--$0.92\,\micron$). 

\subsubsection{EXT- like images}
\label{sec:data-img-ext}

The EXT-like ground-based data set in the $g$, $r$, $i$, and $z$ bands is composed of coadded images from publicly available data in the COSMOS field, obtained by a Dark Energy Camera (DECam) community program.\footnote{Data available on \url{http://archive1.dm.noao.edu/}; program number: 2013A-0351.} The mosaics were created using the Cosmology Data Management system (\texttt{CosmoDM}, \citealt{Mohr2012,Desai2012,Desai2015,Hennig2017}).

The data processing and calibration follow the standard procedure, as outlined in \citet{Hennig2017}, where the single-epoch images are astrometrically calibrated to 2MASS \citep{Skrutskie2006} and internally photometrically calibrated using stellar sources that are common between pairs of overlapping images. Masking of transient artifacts is applied using the method described in \citet{Desai2016}. The images are coadded and resampled onto the \textit{Euclid} pixel grid ($ \ang{;;0.1}$/px).
Table~\ref{tab:extLike} lists the properties of the DECam coadded images prepared for this work. 

\subsubsection{NIR-like images}
\label{sec:data-img-nir}

The NIR images ($Y$, $J$, and $H$) were produced by Terapix as part of the UltraVISTA release 1\footnote{\url{http://www.eso.org/sci/observing/phase3/data_releases/ultravista_dr1.html}} \citep{McCracken2012}, and were resampled onto the \textit{Euclid} pixel grid. These images have similar depths to those quoted in the \textit{Euclid} Red Book \citep{Laureijs2011}, so it was decided not to add any additional noise. It must be noted, however, that the $Y$, $J$, and $H$ filters differ significantly from the equivalent \textit{Euclid} filters since the \textit{Euclid} ones are designed to leave no gap between the filters and the \textit{Euclid} $H$ band extends up to $2\,\micron$.
Table~\ref{tab:extLike} shows the properties of the NIR-like coadded images. More details on the UltraVISTA images can be found in \citet{McCracken2012}.


\subsection{Photometry}
\label{sec:data-phot}

Source detection was performed on the $VIS$-like image. The PSF of all images were homogenized to the $g$-band one, which has the poorest resolution among the eight images. The fluxes were extracted from the images using \texttt{SExtractor} 2.19.5 \citep{Bertin1996} in dual-image mode. Total flux measurements were performed on the $VIS$-like image from \texttt{SExtractor} \texttt{FLUX\_AUTO} counts. Fluxes in the other bands were measured in apertures on PSF-matched images. The conversion between counts in band $X$, $C_{X}$, as measured by \texttt{SExtractor}, and fluxes, $F_{X}$, in $\muup\rm{Jy}$ was performed using
\begin{equation}
F_{X} = C_{X} \,10^{0.4(23.9-\mathrm{ZP}_{X})}.
\end{equation}
The zero-points (ZP$_{X}$) of band $X$ can be found in Table~\ref{tab:extLike}.

Aperture fluxes on the PSF-matched images were computed in circular apertures of $n$ times the flux profile full width at half maximum (FWHM) of the PSF in the $g$-band image. The flux in each band was scaled to total flux according to the following equation:
\begin{equation}
F_{ X, \mathrm{tot}} = \left ( \frac{F_{ X,\mathrm{aper}}}{F_{ VIS, \mathrm{aper}}} \right )  F_{ VIS, \mathrm{tot}},
\end{equation}
where $F_{ X,\mathrm{aper}}$ is the measured aperture flux in band $X$ for which the PSF has been matched to the one of the $g$-band, $F_{VIS, \mathrm{aper}}$ is the aperture flux in the $VIS$-like-band with PSF matched to the $g$-band one, and $F_{VIS, \mathrm{tot}}$ is the total flux extracted in the $VIS$-like-band with its original PSF. Fluxes were obtained for three different apertures sizes, with $n=1,2,$ and $3$ times the FWHM of the {\em g}-band PSF. Flux uncertainties were also scaled on the basis of the ratio between the total $VIS$-like flux and the $VIS$-like PSF-matched one measured in an aperture:
\begin{equation}
    \begin{aligned}
F_{ \mathrm{err},X, \mathrm{tot}} = (F_{ \mathrm{err},X,\mathrm{aper}}) \frac{F_{ VIS, \mathrm{tot}}}{(F_{ VIS, \mathrm{aper}})},
    \end{aligned}
\end{equation}
where $F_{ \mathrm{err},X,\mathrm{aper}}$ is the aperture flux error in band $X$.

To take into account pixel correlations coming from the resampling, which underestimate the measured flux errors, the errors were corrected according to the difference measured in $2\arcsec$ diameter apertures between the sky background noise and the mean variance computed from the weight maps. The corrections (multiplicative factors on flux errors) are given in Table~\ref{tab:extLike}.


\subsection{Catalog}
\label{sec:data-cat}

All objects detected by \texttt{SExtractor} are included in the catalog. Objects with problematic photometry (mostly located at the borders of masked regions), bad \texttt{SExtractor} flags, or with zero weight in at least in one of the weight maps (mostly objects outside the near-IR footprint) have been flagged. Galactic extinction correction factors for the fluxes in all bands, derived from \citet{Schlegel1998}, are also provided for all sources but were not applied directly to the extracted photometry.   

The catalog is divided into two regions based on right ascension $\alpha$, defining two sub-catalogs: the calibration catalog, with $\alpha>\ang{150.125}$; and the validation catalog, with $\alpha\leq \ang{150.125}$. The first catalog is used for the calibration of the different methods to be tested, and the second one is used to assess the performance of all the codes. The number of sources is $198\,435$ in the calibration sample and $192\,864$ in the validation sample. 


Both photometric catalogs have been matched to the master spectroscopic catalog maintained by M. Salvato, which is available within the COSMOS collaboration and contains approximately 50\,000 objects (including around 30\,000 with high-confidence flags), which serves as our primary reference to measure photo-$z$ performance. Only the spec-$z$s for the calibration sample are provided as part of the challenge.

\begin{figure}
    \centering
    \includegraphics[width=\linewidth]{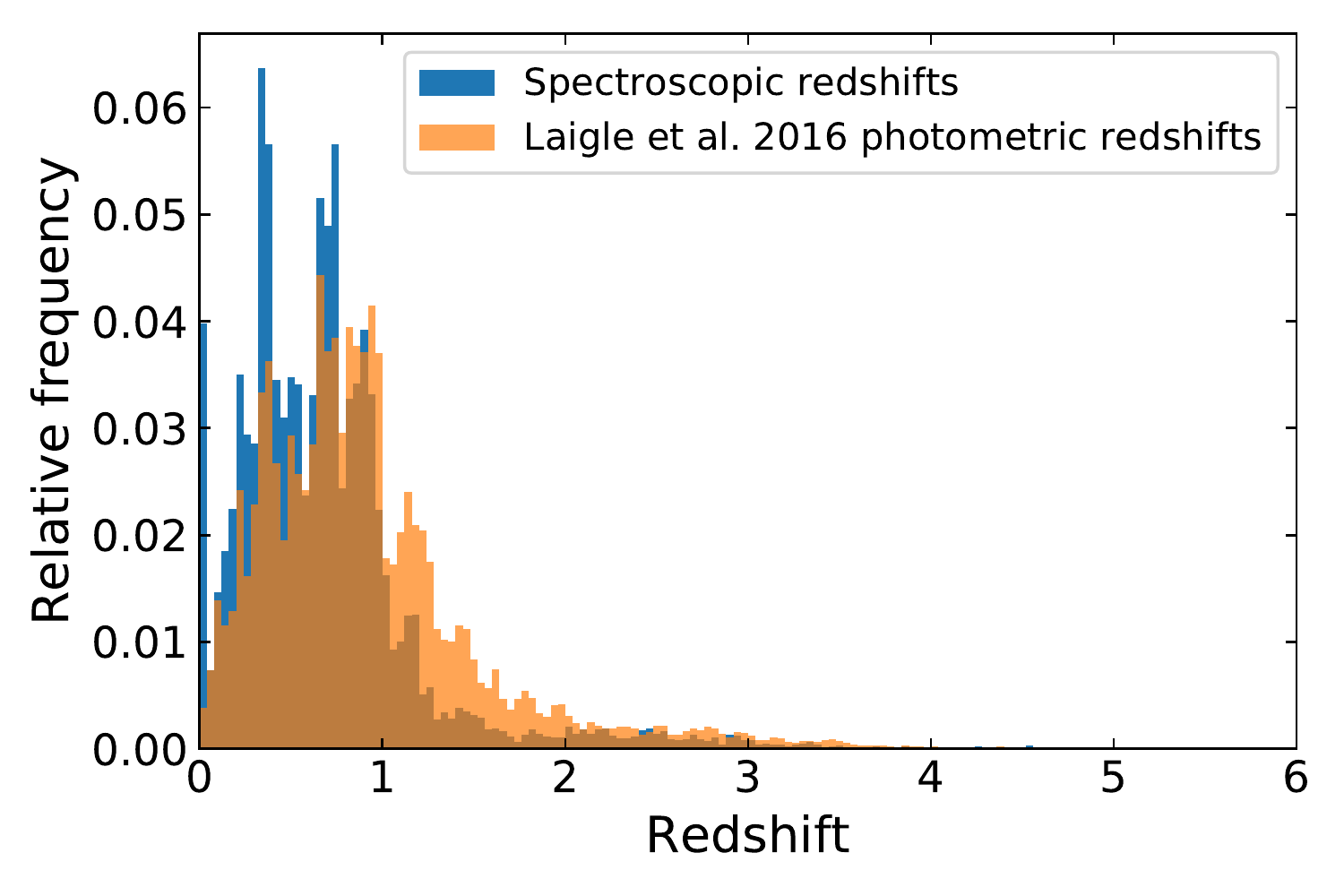}
    \caption{Distribution of the full sample of reliable redshifts. Spectroscopic-redshifts are in blue (27\,872 sources), while \citet{Laigle2016} 30-band photo-$z$s are in orange (107\,267 sources).}
    \label{fig:field_spectro}
\end{figure}

In addition to spectroscopic-redshifts, we matched the photometric catalog with the highly reliable photo-$z$s from \citeauthor{Laigle2016} (\citeyear{Laigle2016}, hereafter L15)  that have been obtained using deep, 30-band photometry (scatter $\sigma = 0.01$ and outlier fraction $\eta=1.7\%$ for $22<i_{AB}<23$ sources). Figure~\ref{fig:field_spectro} compares the distribution in redshift of the spec-$z$s and of the L15 photo-$z$s. The 30-band photo-$z$s are also included in the calibration sample and can be used to calibrate the different methods.

Stars and active galactic nuclei (AGN) in the field are separated from the galaxies by matching our catalog to the point source catalog from \cite{Leauthaud2007}, the L15 catalog, and the catalog of X-ray detected AGN from \citet{Marchesi2016}. Objects are classified as stellar if they are flagged as such in \citet{Leauthaud2007}, or when the spec-$z$s or the L15 photo-$z$s are consistent with 0 and the \texttt{SExtractor FLUX\_RADIUS\_DETECT} measurements are smaller than 1.5\,pixel. Sources with X-ray detections are flagged as AGNs.

\subsection{\textit{Euclid} shear sample}
\label{sec:data-shear}
Unless specified, we focus on the sources in the \textit{Euclid} shear sample. This sample is defined by the set of galaxies with a detection in the $VIS$-like band with signal-to-noise ratio $\rm{S/N}>10$, with $m_{\rm{VIS}}~<~24.5$, that are not flagged as having poor photometry, and that are not flagged as AGNs. In addition, galaxies in the \textit{Euclid} shear sample must have a photo-$z$ in the range $0.2<z_{\rm phot} \leq2.6$, meaning that the detailed composition of this sample is method dependent.


\section{Methods}
\label{sec:methods}

Thirteen different methods have been tested on the data set (see Table~\ref{tab:methods} for a summary). In this section we provide brief descriptions of the algorithms and of the configurations that were used. A requirement of the challenge is that all methods should provide photo-$z$ point estimates (a single value representative of the PDZ, e.g., the mean, the median, the mode, etc.), a probability distribution of the redshift (PDZ), as well as a usability flag (USE flag equals to 0 or 1) for all sources in the validation catalog. This usability flag, indicating whether the participant considers the photo-$z$ estimate reliable or not, is defined freely by the participants. The results for rejected objects with $\rm{USE}=0$ are not accounted in the computation of the metrics. In the spirit of the challenge, the choices for the configuration of the different methods using the calibration catalog data were made independently by the subgroups of authors that ran the codes.

\begin{table*}[h]   
    \centering
    \caption{Summary of the different methods compared in this work, including the name of the code, the type of the approach (template-fitting or machine-learning) and whether a rejection of the results is applied or not. We qualify as strong a rejection of more than 15\% of the full spectroscopic sample (more than 10594 sources remaining; see Table~\ref{tab:statsAll}), otherwise it is considered to be a weak one. }
    \begin{tabular}{llll}
    \hline\hline
     & Type & Rejection &   \\
    \hline
    \rule{0pt}{1.2em}\texttt{Le Phare} & Template-fitting & Weak \\
    \texttt{CPz} & Random forest classification + template-fitting & Weak \\
    \texttt{Phosphoros} & Template-fitting & No  \\
    \texttt{EAzY} & Template-fitting & Strong  \\
    \texttt{METAPHOR} & Machine-learning: neural network & Strong   \\
    \texttt{ANNz} & Machine-learning: neural network & No  \\
    \texttt{GPz} & Machine-learning: Gaussian processes &  Weak\\
    \texttt{GBRT} & Machine-learning: boosted decision trees & Weak  \\
    \texttt{RF} & Machine-learning: random forest & No  \\
    \texttt{Adaboost} & Machine-learning: boosted decision trees & No  \\
    \texttt{DNF} & Machine-learning: nearest neighbor &  Strong  \\
    \texttt{frankenz} & Machine-learning: nearest neighbor & Strong  \\
    \texttt{NNPZ} & Machine-learning: nearest neighbor & No   \\
    \hline
    \end{tabular}
    \label{tab:methods}
\end{table*}

\subsection{\texttt{Le Phare}}
\label{sec:methods-lephare}

\texttt{Le Phare} (\citealt{Arnouts2002,Ilbert2006}) is a template-fitting method. The photo-$z$s for this work were derived following the recipes outlined in \citet{Ilbert2009}. Thirty-three spectral energy distribution (SED) templates were used: the 31 COSMOS templates \citep{Ilbert2009}, that includes elliptical and spiral galaxies from the \citet{Polletta2007} library (some of them being linearly interpolated to refine the sampling in color-redshift space) and young and blue star-forming galaxies whose templates were generated with \citet{Bruzual2003} stellar population synthesis models; and two templates of elliptical galaxies generated with an exponentially decaying star-formation history (SFH; following \citealt{Ilbert2013}). Extinction was added as a free parameter ($E_{B-V}<0.5$) on templates of type Sc and bluer. The \citet{Calzetti2000} attenuation curves were considered, adding a possible bump at 2175\AA, as well as the \citet{Prevot1984} attenuation curve. Emission lines were added to the templates using an empirical relation between UV light and emission line fluxes \citep{Ilbert2009}. Line fluxes were allowed to vary by a factor 2, but without changing the emission line ratios.

In the fit of the 2-FWHM photometry, a minimum error of 0.01 mag for all the visible bands was applied, and a minimum error of 0.03 mag for all the NIR \textit{Euclid} bands was applied. A cut in absolute magnitude was applied, discarding all solutions with galaxies brighter than $M_g=-24$. An optimization of the zero-points was made using the method of \citet{Ilbert2006}. Offsets as large as 0.07 mag were applied to two bands, namely the $i$ and $Y$ bands.

Redshift point estimates are given as the median of the marginalized PDZ. 
All sources with a 68\% confidence intervals around the median larger than $0.3(1+z)$ were flagged with $\rm{USE}=0$.

\subsection{\texttt{CPz}}

Classification-aided photometric-redshift estimation (\texttt{CPz}; \citealt{Fotopoulou2018}) is a hybrid approach to compute photometric redshifts. This methods uses random forest \citep{Breiman2001} to assign each object its optimal class, and then uses traditional SED fitting \citep[Le Phare;][]{Arnouts1999} for the photometric-redshift estimations. The goal of this method is to use a restricted library of templates optimized for each of the galaxy classes considered, aiming to reduce degeneracies between models. It can be considered as a generalization of the approach described in \citet{Salvato2011}.

The data were split into three equal parts, used for training, validation, and testing, respectively. Three distinct random forest classifiers were trained to assign each object into: \textit{(i)} star versus not star; \textit{(ii)} one of the five galaxy classes (passive, starforming, starburst, AGN, or QSO); and \textit{(iii)} photometric redshift outlier. All models were fit to the data and labels were assigned based on the SEDs that provide the best photometric redshifts. The galaxy models (passive, starforming, starburst) are the 31 COSMOS templates used in \citet{Ilbert2009}, while the AGN and QSO templates are from \citet{Salvato2011}. A detailed description of the model set up can be found in \citet[][Case III]{Fotopoulou2018}. Briefly, models were generated  at $0<z<6$ with $\Delta z=0.01$. Attenuation values $E_{B-V}=0, 0.05, 0.1, 0.15, 0.2,$ and $0.3$ were used. Emission lines were added only for the normal galaxy templates, as the AGN and QSO templates are empirical and already contain emission lines.

The classification was performed in color space, by taking all color combinations of the input photometry without any input redshift information. Once the three classifiers were trained and applied to the entire sample, the redshift solution was assigned using the model library identified by the classifier as optimal. Additionally,  sources classified as stars ($\mathrm{P_{star}\geq0.5}$) or outliers ($\mathrm{P_{outlier}}\geq0.5$) were rejected ($\rm{USE}=0$). Since this application concerns the estimation of photometric redshifts for \textit{Euclid}, sources that are classified as AGN or QSO are also rejected, since they typically have lower quality photo-$z$. 

\subsection{\texttt{Phosphoros}}

\texttt{Phosphoros} (Paltani et al.; in prep.) is a Bayesian template-fitting tool developed with the aim of being run in a computer-intensive processing environment while including most of the advanced features found in similar codes, such as the use of upper limits, zero-point corrections, consideration of emission lines, various intrinsic extinction curves, etc. \texttt{Phosphoros} will implement unique features, like complex user-defined priors (e.g., from luminosity functions), the choice between different intergalactic medium prescriptions, the sampling of the posterior, etc. Because it is still under active development, the only advanced and unique feature that we use here is the improved treatment of Galactic reddening \citep{GalametzA2017}.

The 2-FWHM aperture photometry was selected for the EXT-like and NIR-like bands, and the total flux for the \textit{VIS}-like band. The photometric data were fit with the 31 COSMOS galaxy (SED) templates (see Sect.~\ref{sec:methods-lephare}) with a similar configuration as in \citet{Ilbert2013}, from $z=0.01$ to $z=5.99$ with step size of $\Delta z=0.02$. Intrinsic reddening was set as a free parameter, with $E_{B-V}\leq0.5$ and several extinction laws (\citealt{Prevot1984}, \citealt{Calzetti2000} and modified Calzetti laws including a bump at 2175\,\AA\ as in \citealt{Ilbert2009}). For templates representing galaxies with types earlier than Sc, no extinction was added.  The  H\,$\alpha$ to H\,$\delta$, [O\,{\sc ii}] 3727\,\AA\ and [O\,{\sc iii}] 4959+5007\,\AA\ emission lines were added to all templates using an empirical relation between H\,$\alpha$ and other emission line fluxes, which were recalibrated using line fluxes measured from the Sloan Digital Sky Survey \citep{Thomas2013}. The Milky Way reddening was treated as prescribed in \citet{GalametzA2017} by applying a reddening correction to the templates and fitting uncorrected photometry. Zero-point corrections to the photometric calibration were computed in the same way as in \citet{Ilbert2006} using 2000 randomly selected galaxies with spec-$z$s from the calibration catalog. No luminosity prior has been used.

The PDZs are constructed by marginalizing the likelihood over the template and reddening dimensions. The point estimates used in the rest of the analysis were computed from the mode of the PDZ for each object. Finally, no rejection was made on the quality of the results (USE flag was set to 1 for all sources).

\subsection{EAzY}

The setup for the \texttt{EAzY} code \citep{Brammer2008} was kept close to the default configuration. All template combinations of the seven base SED components with added emission lines were allowed, plus a young, heavily dust-reddened galaxy SED (which is not allowed to be combined with the other SEDs). The extended $r$-band magnitude based prior, $p(z~|~m_{r})$, was applied and a flat $\Lambda$CDM cosmology with $H_0=70~{\rm km\,s^{-1}\,Mpc^{-1}}$ and $\Omega_{\rm m}=0.3$ for luminosity computation was used. The run was performed using the 2-FWHM aperture photometry without the $VIS$-like-band. A single best $\chi^2$ value among the possible template combinations was returned at each redshift, which was then combined with the magnitude-based prior to produce the galaxy PDZs.

Similar to other template-fitting based methods, the \texttt{EAzY} code includes the flexibility to apply corrections to photometric zero-points and a systematic uncertainty in measured photometric fluxes. However, there is also a wavelength-dependent template uncertainty function which is controlled by a parameter that governs its amplitude. These nine parameters (namely seven zero-points, fractional systematic flux error and template error function amplitude) were optimized using the spectroscopic training sample and the Python function \texttt{minimize} from the \texttt{scipy.optimise} package. Values were initialized at zero for the zero-point adjustments, $3\%$ for the systematic flux error, and $0.7$ for the amplitude of the template error function. The loss function is a linear combination of the normalized median absolute deviation, mean point redshift bias, Kullback--Leibler divergence of the histogram of probability integral transform values (see Sect.~\ref{sec:results-PDFs} for more information), and outlier fraction. Each term in the loss function was scaled such that a value of unity represented good performance. The point redshift used for the first two terms, and for the tomographic bin assignment, is \texttt{z\_peak}, the mean redshift of the most probable peak in the PDZ. Finally, objects were flagged as unreliable if their odds value was smaller than 0.91. The odds value quantifies the extent to which a PDZ is single-peaked (see \citealt{Brammer2008}), and this value was chosen as a compromise between sample completeness and performance on the same set of metrics that were used in the loss function.

\subsection{\texttt{METAPHOR}}

\texttt{METAPHOR} (Machine-learning Estimation Tool for Accurate PHOtometric Redshifts; \citealt{Amaro2019,Cavuoti2017}) has a modular workflow, designed to produce the redshift point estimations and the PDZs. Its internal photo-$z$ estimation engine is based on the MLPQNA machine-learning model (Multi Layer Perceptron with Quasi Newton Algorithm; \citealt{brescia2013,cavuoti2015}). 

The key concept of \texttt{METAPHOR} is to perform a series of independent photometry perturbations to take into account the contribution of the uncertainty induced by the photometric errors within the PDZs. In other words, the idea is to obtain an estimation of the photo-$z$ PDZs based on the predictive performance evaluation of the trained MLPQNA model by varying the magnitudes within the photometric errors and considering the distribution of the multiple output as the PDZ. The perturbation method is based on the addition of a variable random Gaussian noise to the photometry and a polynomial fitting of the photometric trend to reproduce the inner distribution of the error.

In practice, each PDZ was based on the following steps: \textit{(i)} training of MLPQNA with unperturbed SEDs (the training set); \textit{(ii)} producing $N$ different instances of any source SED (the blind test set) contaminated by photometric noise; \textit{(iii)} deriving $N+1$ photo-$z$ estimates for the sources with the trained model (i.e., $N$ perturbed + the original one); and \textit{(iv)} binning in photo-$z$ of the $N+1$ values, thus calculating for each one the probability that a given photo-$z$ estimation belongs to each bin (i.e., obtaining the PDZ). In the particular case of this \textit{Euclid} challenge, $N=999$ was used. The point estimate is the value among the $N+1$ values that is the closest to the non-perturbed value.


The training was done with the 2-FWHM photometry in all bands, considering all galaxies (including AGNs) with spec-$z$s and flagged as having proper photometry. In order to introduce a quality flag for the estimates, a two-step analysis was performed. First, a selection on the photometry in which all objects with $\rm{S/N}\leq3$ in any of the $griz$ bands, $\rm{S/N}\leq5$ in any of the $YJH$ and $VIS$-like bands, or a \texttt{SExtractor} detection flag $\geq4$, were marked with the flag $\rm{USE}=0$. Second, a further refinement of the flag assignment was performed through a selection on the PDZ to avoid overly wide PDZs. The criteria were: the maximum value of a PDZ must be $\geq0.09$; the width of its primary peak $\leq0.44$ in redshift; and the overall distribution must be $\leq 2$.

\subsection{\texttt{ANNz}}

\texttt{ANNz} \citep{Collister2004} is a neural-network-based photometric redshift code that uses a training set with both photometric and spectroscopic information to learn the mapping between the color-magnitude space of galaxies to their redshifts. The learning algorithm minimizes the mean-squared error between predicted and (assumed to be) true, spectroscopic-redshifts.
The learned function is then an estimate of the mean of the conditional distribution $p(z | \mathbf{f})$, where $z$ denotes the redshift and $\mathbf{f}$ the vector of galaxy colors and their magnitudes. The learned model is then applied to the full data sample to obtain photometric redshift estimates. 

To obtain error bars on these point estimates, one can subsequently train an additional neural network to predict the mean-squared error between true (i.e., spectroscopic) redshift and the predicted redshift from the previously trained model. This is done using the same basic setup, meaning, again by minimizing the mean-squared error. The resulting predictions from this second run then provide an estimate for the variance of the conditional distribution $p(z | \mathbf{f})$. These error bars can only be expected to be well calibrated if enough training data are available, the training data are representative, and the conditional distributions $p(z | \mathbf{f})$ are close to Gaussian. The interested reader is referred to \citet{Rau2015} for a discussion of the impact of these distributional assumptions on photo-$z$ results.

The calibration sample was split into two representative subsamples. The first one was used for training, and the second one was used for testing the models. 

\subsection{\texttt{GPz}}

\texttt{GPz} is a machine-learning tool that models the relation between input data (e.g., observed magnitudes, which we call ``color'' for simplicity) and an output value (the redshift). The model used by \texttt{GPz} is a linear combination of multivariate Gaussian functions (called ``basis functions''; here 100 are used). In addition to learning the mean relation between colors and redshifts, \texttt{GPz} also learns the scatter of the redshift at a given position in color space, as well as the density of the training data. It uses this information, together with knowledge of the uncertainties on the observed colors, to make a prediction of the PDZ. At a given position in color space, the predicted PDZ will be broader if: \textit{(i)} the colors are uncertain; \textit{(ii)} the range of matching spec-$z$ is large; or \textit{(iii)} there is a lack of training data. A limitation of this model is that the distribution is forced to be Gaussian  \citep{Almosallam2016a,Almosallam2016b}. All the predictions here are produced with the C++ version of \texttt{GPz}, available in the \textit{Euclid} GitLab as ``\texttt{PHZ\_GPz}''. Here, the model was trained on the shear sample, since this is the sample for which the metrics need to be optimized in \textit{Euclid}. The 2-FWHM fluxes were used, and the fluxes were converted to ``luptitudes'' \citep{Lupton1999} before prediction and training.  


First, \texttt{GPz} models the distribution in color space of the validation data set (the one for which we want to make predictions) using Gaussian mixture models. It then applies this Gaussian mixture to the training set to: \textit{(i)} weight the training data so its color distributions match the validation data; and \textit{(ii)} split the color space in several (typically 5) distinct regions in which separate \texttt{GPz} models will be trained. The first point deals with any potential bias in the color distribution of the training set, while the second point effectively increases the number of basis functions used to model a given region of color space without paying for the full computational costs. 


\subsection{\texttt{Gradient boosted regression trees}}

Gradient boosted regression trees (\texttt{GBRT}) is a machine-learning method based on the \texttt{sci-kit learn} gradient boosted decision tree algorithm \citep{Friedman2001,Pedregosa2011}. For its training, galaxies and AGNs with good quality spectroscopic or 30-band photometric redshifts from L15 and detected in at least four bands were selected from the calibration catalog, leading to a training sample of around $ 1.4 \times 10^{5}$ sources. This sample size is increased by an order of magnitude by synthesizing 10 brighter and fainter versions of each source. The 2-FWHM aperture photometry in the \textit{g}, \textit{r}, \textit{i}, and \textit{z} bands, the 1-FWHM aperture photometry in the \textit{Y}, \textit{J}, and \textit{H} bands, and the \textit{VIS} total photometry were used to train the algorithm. 

A point estimate was determined for each source, and a PDZ was constructed through processing of 1000 realizations perturbed by a Gaussian error for each source. \texttt{GBRT} provides an indication of the most useful bands for the photo-$z$ determination, those being the $g$, $Y$, $J$, $H$, and $VIS$-like bands. Sources located in regions of this color space that were not covered by sources from the training sample were rejected ($\rm{USE}=0$).

\subsection{\texttt{Primal Random Forest}}
\label{sec:methods-RF}

The Primal Random Forest (\texttt{RF}) is based on the \texttt{sci-kit learn} random forest regressor  \citep{Breiman2001,Pedregosa2011} wrapped in the Primal framework.\footnote{\url{https://github.com/andreatramacere/primal}} The training was done by selecting all sources with reliable spectroscopic-redshifts in the calibration catalog. The features used were the 2-FWHM aperture fluxes in all standard bands and the total fluxes in the $VIS$-like band, along with the flux ratios and flux errors. The calibration sample was split into training (20\%) and testing (80\%) sets using a reshuffling procedure with stratified sampling to insure that both sets were representative of the full sample. \texttt{RF} is optimized by performing a recursive feature elimination, selecting the most important features that provide the minimum outlier fraction. 

The validation set was processed 5000 times with perturbed fluxes according to their errors. The PDZs were constructed by binning the 5000 results for each source. The point estimates are the modes of the constructed PDZs. No rejection was made on the quality of the results, so that the USE flag is set to 1 for all objects with good photometric flags.

\subsection{\texttt{Primal Adaboost}}

The \texttt{Primal Adaboost} method is the \texttt{sci-kit learn} Adaboost regressor algorithm \citep{Freund1997} wrapped in the Primal framework. We used boosted decision tree regressors. The training and the processing were done in the exact same way as for the RF, which is described in Sect.~\ref{sec:methods-RF}.

\subsection{\texttt{DNF}}

\texttt{DNF} (Directional Neighborhood Fitting; \citealt{de2016dnf}) computes the photo-$z$ of a galaxy by a linear combination of multi-band fluxes. The coefficients of the prediction hyperplane are determined by fitting the equation with a subsample of neighbors within a reference sample whose spectroscopic-redshifts are known.  A novel metric (``directional neighborhood'') is defined to account simultaneously for the magnitudes and colors of the galaxies. 
The PDZs are computed from the residuals of the fit and reflect the uncertainties and degeneracies associated with individual photo-$z$ predictions (see details in \citealt{de2016dnf}). \texttt{DNF} also produces a second photo-$z$ ($z_{\rm{phot,2}}$ estimate) as the redshift of the nearest directional neighbor in the reference sample. The stacking of $z_{\rm{phot,2}}$ values for the whole sample provides a reference redshift distribution estimation, if the target galaxies are well represented within the training sample.


\texttt{DNF} was run on galaxies only, using the 3-FWHM photometry. \texttt{DNF} provides an error estimation of individual photo-$z$s that accounts for flux uncertainty, also tagging the lack of neighbor reference samples. This parameter allows one to cut the samples according to different precision, bias, or completeness requirements. In this test, precision was prioritized over bias and completeness, producing an aggressive cut of 50\% of the sample. Other configurations are possible such as those focusing only on removing the most unreliable photo-$z$s.

\begin{figure*}
    \centering
    \includegraphics[width=\linewidth]{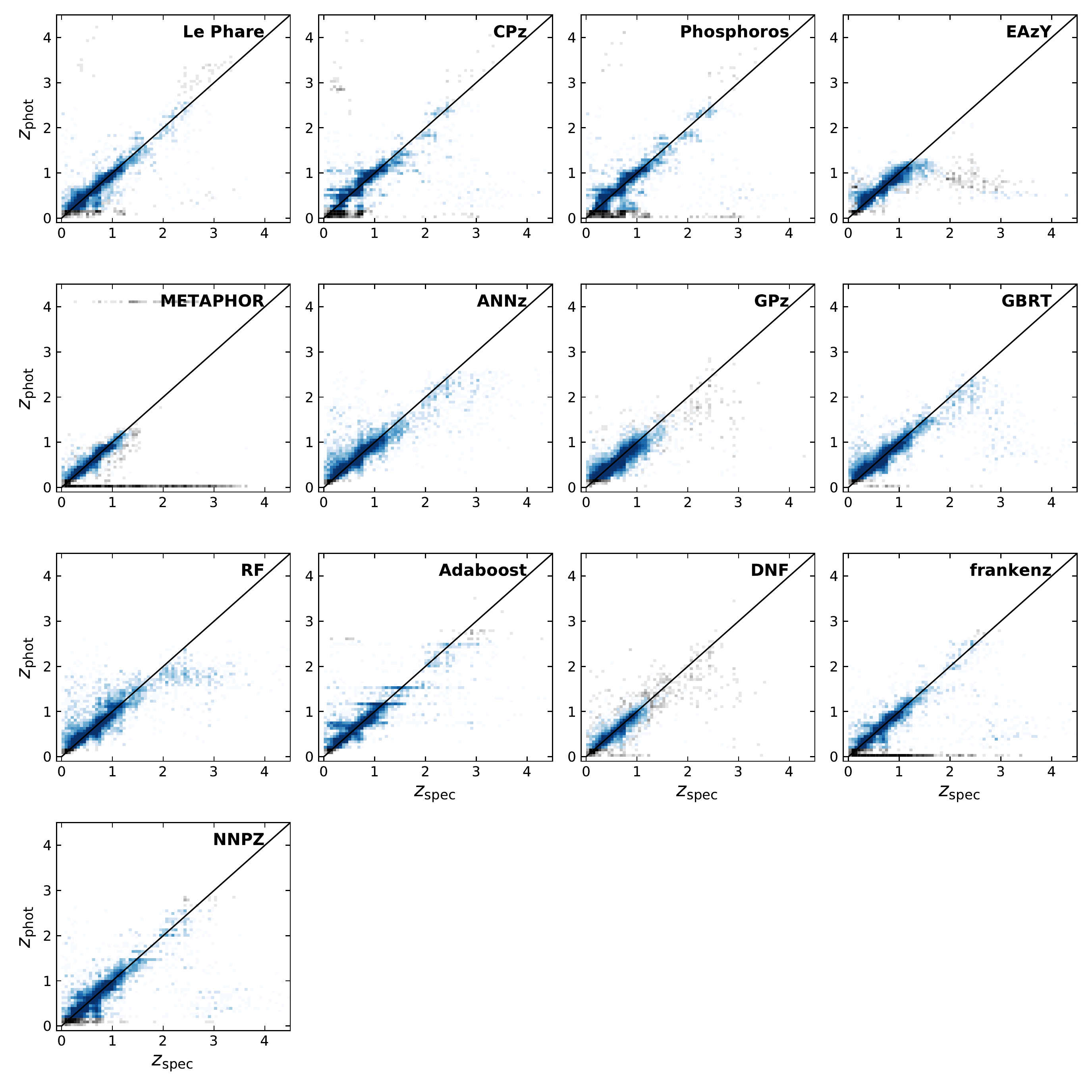}\\
    \includegraphics[width=\linewidth]{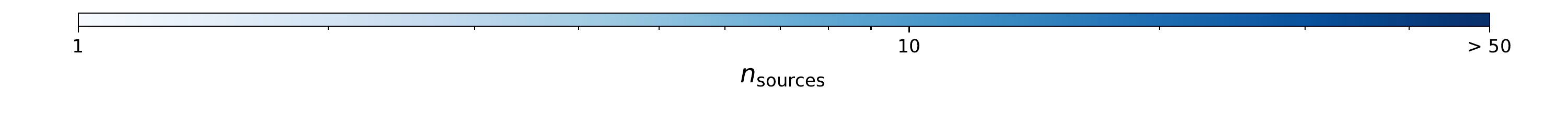}
    \caption{Density maps of photo-$z$ versus spec-$z$ for all the tested methods: blue are sources within the \textit{Euclid} sample; gray are sources outside of the \textit{Euclid} sample. The statistics on the photo-$z$s are presented in Fig.~\ref{fig:PEstats} and listed in Tables~\ref{tab:statsAll} and~\ref{tab:statsEuclid}. Undefined or negative photo-$z$s have been set to 0, explaining the presence of horizontal lines in some panels (e.g., \texttt{METAPHOR} and \texttt{frankenz}).}
    \label{fig:densityPlot}
\end{figure*}

\subsection{\texttt{frankenz}}

\texttt{frankenz} (\citealt{Tanaka2018}; Speagle et al., in prep.) adopts a Bayesian-oriented nearest-neighbors-based approach 
that attempts to properly account for measurement errors within both training and testing sets 
when making photo-$z$ predictions. Neighbors were selected using a Monte Carlo approach over repeated
realizations of the photometric errors, after which priors over the training set (here assumed to be uniform)
and the likelihoods between each unique training-testing object pair were computed explicitly in flux space.
PDZs were then constructed using a posterior-weighted average of each object's redshift kernel.
Objects with large best-fit reduced $\chi^{2}$ values among the set of nearest neighbors were flagged not to be used ($\rm{USE}=0$).

\subsection{\texttt{NNPZ}}

\texttt{NNPZ} (Nearest-Neighbor Photometric Redshift) is a machine-learning algorithm that consists in a $k$-nearest neighbor method in flux space, developed by J. Coupon, that is designed to produce PDZs and was applied to the HSC-SSP survey \citep{Tanaka2018}.

An improved version of the algorithm was used here, which takes into account errors when searching for the neighbors and weights them according to some distance definition. For efficiency, in this implementation of \texttt{NNPZ} the process is split into three stages. First, \texttt{NNPZ} reduces the search space by selecting a candidate set of neighbors using a \emph{k-dimensional tree} and Euclidean distances, which allows for look-ups in $\mathcal{O}(\log{} n)$ steps.
Over this initial candidate set, the final neighbors are searched using a $\chi^2$ distance, which takes into account both the errors of the reference and the target object. Finally, the weights are computed using the likelihood of the $\chi^2$.

The training was done using the Galactic-reddening corrected 2-FWHM aperture photometry of the sources that were not flagged as stars or AGN. The labels were the reliable L15 photo-$z$s, restricted to $ 0 < z \le 6 $. For the first stage, \texttt{NNPZ} selected 2000 neighbors using the Euclidean distance, then later reduced their number to 30 using the $\chi^2$ distance. The PDZs were constructed by combining the L15 PDZs of the weighted neighbors.

The point estimate is the mode of the PDZs for each source. No rejection was made on the quality of the results, so that the USE flag was set to 1 for all objects with good photometry flags.


\section{Results}
\label{sec:results}

In the following, we consider the \textit{Euclid} shear sample (see Sect.~\ref{sec:data-shear}). In the \textit{Euclid} context we focus on the performance of the different methods in the $0.2<z<2.6$ photo-$z$ range and for source with $\rm{USE}=1$.  In the rest of the analysis, we refer to this selection as the \textit{Euclid} selection. We point out that the \textit{Euclid} selection is different for each method, since each method assigns different photo-$z$s and has different flagging schemes.

\subsection{Point estimates}
\label{sec:results-pointEstimates}

\begin{figure*}
    \centering
    \includegraphics[width=\linewidth]{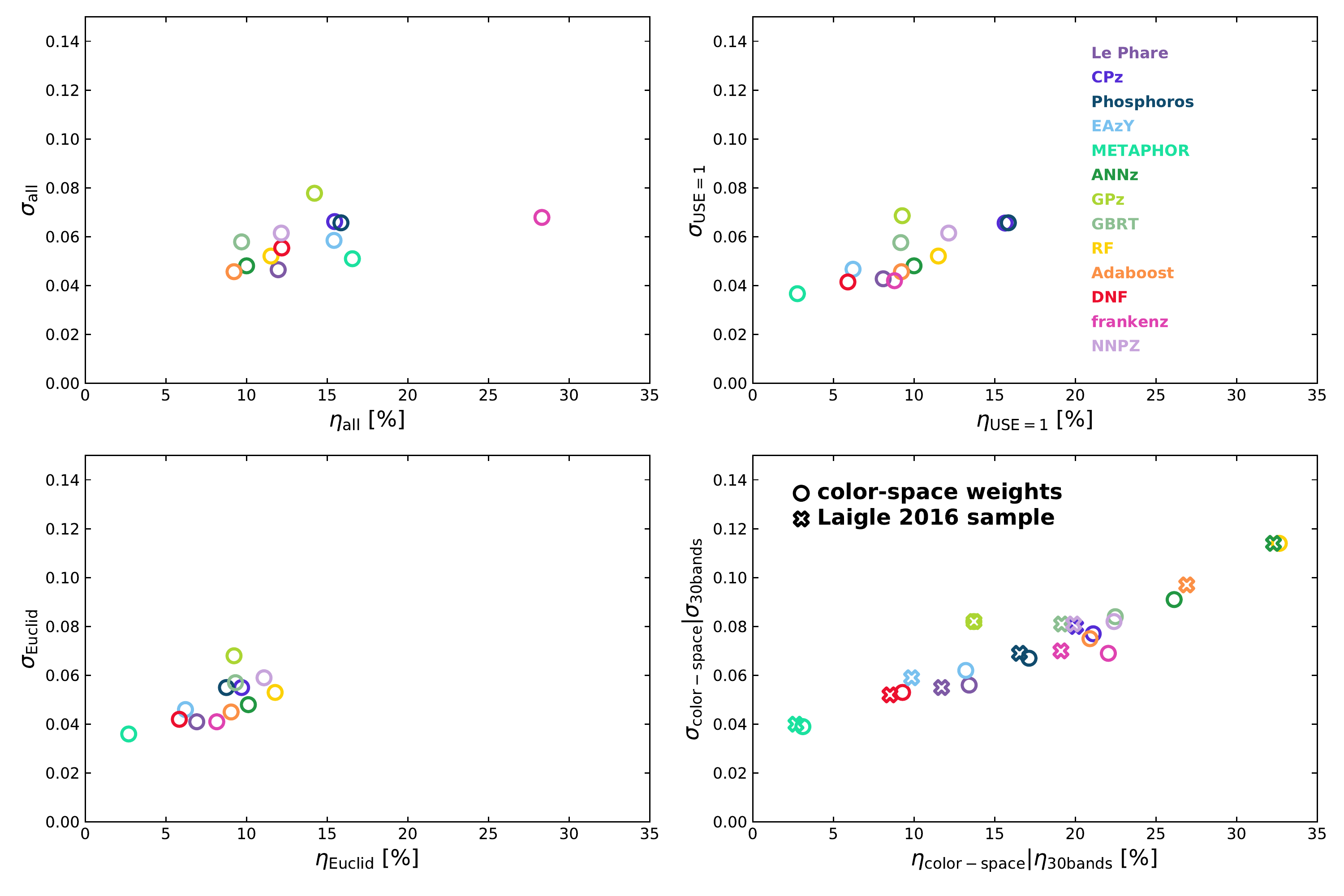}
    \caption{Point estimates metrics results comparison for all the methods. Circles represent the spectroscopic sample and crosses are the L15 one. \textit{Top left}: scatter ($\sigma$) versus outlier fraction ($\eta$) of all methods for the whole spectroscopic sample (12\,463 sources). \textit{Top right}: $\sigma_{\rm{USE=1}}$ versus $\eta_{\rm{USE=1}}$ for $\rm{USE}=1$ selected sources for each method (see Table~\ref{tab:statsAll} for all the values). \textit{Bottom left}: $\sigma_{\rm{Euclid}}$ versus $\eta_{\rm{Euclid}}$ for the \textit{Euclid} selection (see Table~\ref{tab:statsEuclid}). \textit{Bottom right}: $\sigma_{\rm{30bands}}$ versus $\eta_{\rm{30bands}}$ using the L15 photo-$z$ as reference redshift plotted as crosses (see Table~\ref{tab:stats30bands} for all the values) and \textit{Euclid} sample results  weighted with the color-space weights to match the spectroscopic sample to the photometric one plotted as circles (see Table~\ref{tab:statsWeight}). \texttt{RF} values are outside the limits of the plot for the L15 sample due to a large outlier fraction.}
    \label{fig:PEstats}
\end{figure*}

First, we look at the point estimates and assess the quality of the results through the following commonly used metrics: the normalized median absolute deviation of the residuals $\sigma~=~1.4826~\times~\mathrm{median}(|~\Delta z~-~\mathrm{median}(\Delta z)~|)$, \\where $\Delta z=~(z_{\rm spec}~-~z_{\rm phot})/(1~+~z_{\rm spec})$ is the scaled residual between the photo-$z$ and the reference redshift; and the fraction $\eta$ of outlier sources for which $|\Delta z| > 0.15$.

Figure~\ref{fig:densityPlot} shows the density map of the photo-$z$ point estimates versus the spec-$z$s for all thirteen methods. The same plots using the 30-bands photo-$z$s as reference redshift can be found in Appendix~\ref{sec:30bandsPointEstimates}, Fig.~\ref{fig:densityPlot30bands}. All sources without photo-$z$s are set to $z_{\rm{phot}}=0$, which explains the horizontal lines in some of the plots, and are treated as outliers in the computation of the metrics. \texttt{METAPHOR} shows a systematic photo-$z$ at $z=4.12$, which corresponds to the highest redshift in the training sample they considered. The statistics associated with the plots are presented in a graphical form in Fig.~\ref{fig:PEstats}, and all the values are provided in tables in Appendix~\ref{sec:pointEstimateTables}. We note some results that appear similar in Fig.~\ref{fig:densityPlot}, like those of \texttt{Phosphoros} and \texttt{CPz}; this is due to the similarity of the approaches (template-fitting) and configuration (31 COSMOS templates from \citealt{Ilbert2013}), even if the codes are different. On the other hand, the difference in the results between \texttt{Le Phare} and \texttt{CPz} can be explained by the differences in the definition of the point estimate, being the median of the PDZ for \texttt{Le Phare} and the mode for \texttt{CPz}, even if \texttt{CPz} uses \texttt{Le Phare} for the fitting of the templates. Further tests have shown that when run in identical configuration, template-fitting methods provide identical results. This means that the differences observed in the results are not due to differences in performance of the template-fitting methods, but rather to variations in their configurations.

Figure~\ref{fig:PEstats} shows the metrics associated with different reference redshifts and selections applied to the sources. In the top left panel, $\sigma_{\rm{all}}$ and $\eta_{\rm{all}}$ are plotted against each other for the total spectroscopic sample (12\,463 sources with highly reliable spec-$z$ measurements). With its large outlier fraction, \texttt{frankenz} differs greatly from the rest of the methods in the plot. This is due to the sources for which no photo-$z$ are provided, visible in Fig.~\ref{fig:densityPlot} with $z_{\rm phot}=0$. Machine-learning methods seem generally to perform better than the template-fitting ones, especially \texttt{Adaboost} or \texttt{ANNz}. The top right panel of Fig.~\ref{fig:PEstats} presents metrics for the spectroscopic sample, but only considering sources with USE flag equal to 1. In this case, we see some improvement in the results of the methods that apply rejection of the sources for which the predictions are considered less reliable. This phenomenon is particularly obvious for \texttt{METAPHOR}, which shows the best results after this rejection. This demonstrates that the USE flags are able to correctly identify a good fraction of the incorrect predictions, and that they enhance the precision of the results, at the expense of completeness.

For the \textit{Euclid} selection, $\sigma_{\mathrm{Euclid}}$ and $\eta_{\mathrm{Euclid}}$ are presented in Fig.~\ref{fig:PEstats} (bottom left panel). In this range of redshifts, the results are better for all the methods. \texttt{Phosphoros} and \texttt{CPz} show great improvements, since the selection removes low photo-$z$ sources that are poorly constrained due to the absence of $u$-band fluxes in the data. Here again, \texttt{METAPHOR} presents the best values for these metrics.

In order to take into account the fact that the spec-$z$ sample is not representative of the color space of all galaxies, we follow the approach of \citet{Lima2008}. We assign weights to the spectroscopic sample depending on the distances of the 100 nearest neighbors each object has in the color-mag$_{\rm{VIS}}$ space of the full shear sample using a nearest-neighbor method. We compute the indicators $\sigma_{\rm{color-space}}$ and $\eta_{\rm{color-space}}$ with these weights, presented in the bottom right panel of Fig.~\ref{fig:PEstats}. Both scatter and outlier fractions become poorer for most of the methods as one would expect, with the exception of \texttt{METAPHOR} that shows only a small reduction in performance. Summing the weights of the sources in the selection of each method ($N_{\rm{color-space}}$ in Table~\ref{tab:statsWeight}) and comparing this sum to the sum of the weights for all the sources that should be in the \textit{Euclid} sample (9384.4) gives an estimate of the fraction of sources kept by the methods for the photometric sample. For \texttt{METAPHOR} the ratio between the two values is $1/3$, and the ratio is 1/4 for \texttt{DNF} (the median value for all the methods is $\sim0.86$), meaning that the majority of the sources is rejected in the photometric sample in order to keep the precision of the photo-$z$s at the level of the spectroscopic sample.

Another estimate of the quality of the photo-$z$s over the full color space can be obtained by comparing our photo-$z$s with the 30-band ones of \citet{Laigle2016}. The underlying assumption is that the latter photo-$z$s are much more precise than those computed here, thanks to the much deeper and better sampled photometric data. The bottom right panel of Fig.~\ref{fig:PEstats} shows the scatter ($\sigma_{\rm{30bands}}$) and outlier fraction ($\eta_{\rm{30bands}}$) for the \textit{Euclid} selection, computed with L15 photo-$z$s as reference redshifts (see Table~\ref{tab:stats30bands} for all the values). We note that the results with the L15 photo-$z$s are comparable to the color-space-weighted ones. The good match between color-space-weighted results and L15 allows us to consider either of these methods to be good approximations of the scatter and outlier fraction of the photo-$z$ methods over the full photometric sample. 
In the following, we use both the weighted spectroscopic sample and the L15 sample, since we want to assess the quality of the results over the whole color space. Using both samples allows us to consider different systematics in the comparison: the weighted spec-$z$ sample has more reliable reference redshifts, but might not represent the full photometric sample, since some part of the color space might not be covered at all; and the L15 sample, while complete in color space, contains less precise redshifts, as well as some catastrophic failures, because it is based on 30-band photo-$z$s. Methods trained on the spectroscopic sample can be expected to perform better on the weighted spec-$z$ sample, while methods training on L15 data (\texttt{NNPZ} and \texttt{GBRT}), as well as the template-fitting methods, especially if they use the same templates as in L15, might present better results on this sample. 



\subsection{PDZs}
\label{sec:results-PDFs}

Each method provides PDZs for every source. Compared to the point estimates, PDZs include all the information about errors and possible degeneracies of the measurements. We assess here the quality of the PDZs provided by all the methods. We consider only the  \textit{Euclid} sample selection (see Sect. ~\ref{sec:results-pointEstimates}).

\begin{figure}[h!]
    \centering
    \includegraphics[width=\linewidth]{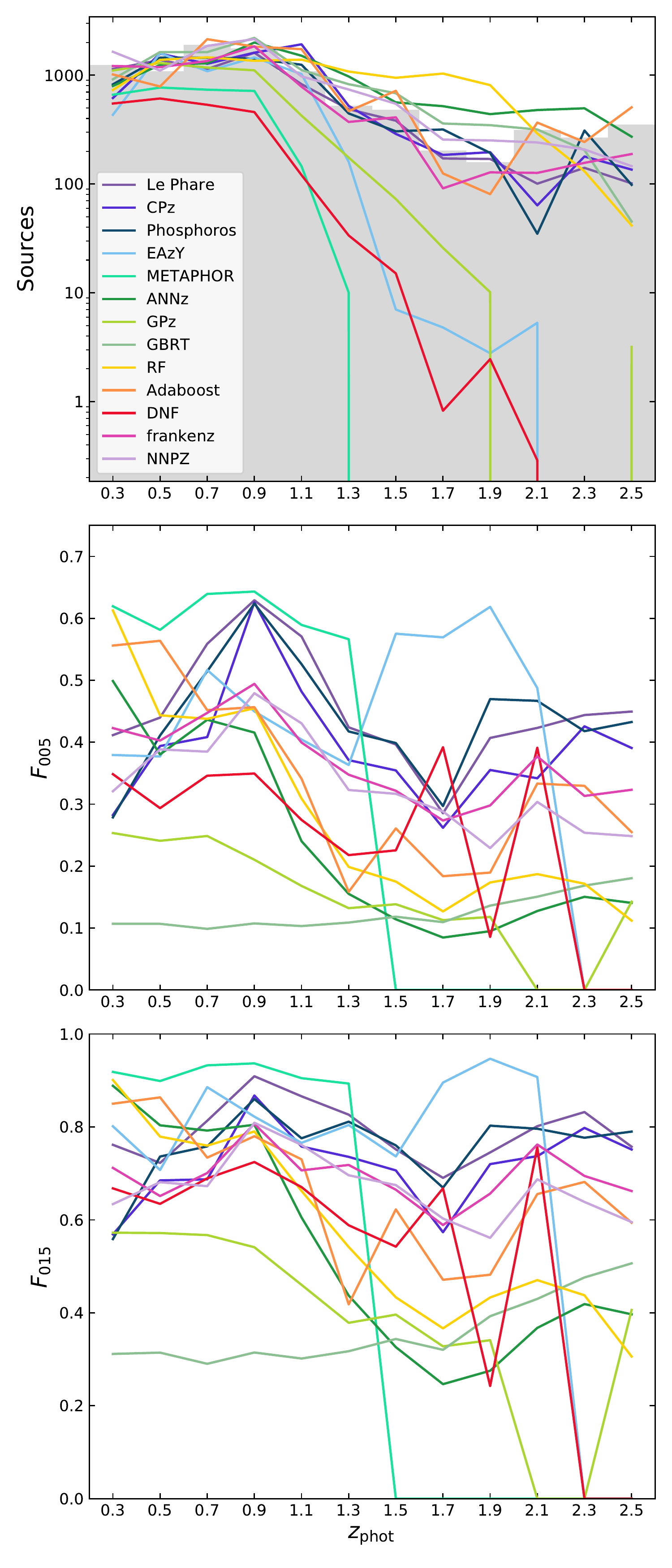}
    \caption{PDZ metrics for the color-space weighted spectroscopic sample. \textit{Top}: Number of sources in the bin. The histogram of the distribution for the sources in the bins according to their spec-$z$s is shown in gray. \textit{Middle}: fraction of the stacked-and-shifted PDZs in $0.05(1+z)$ ($F_{005}$). \textit{Bottom}: fraction of the stacked-and-shifted PDZs in $0.15(1+z)$ ($F_{015}$) for all the tested methods. Fractions close to 1 in a bin indicate good results.}
    \label{fig:specF005F015}
\end{figure}

\begin{figure}
    \centering
    \includegraphics[width=\linewidth]{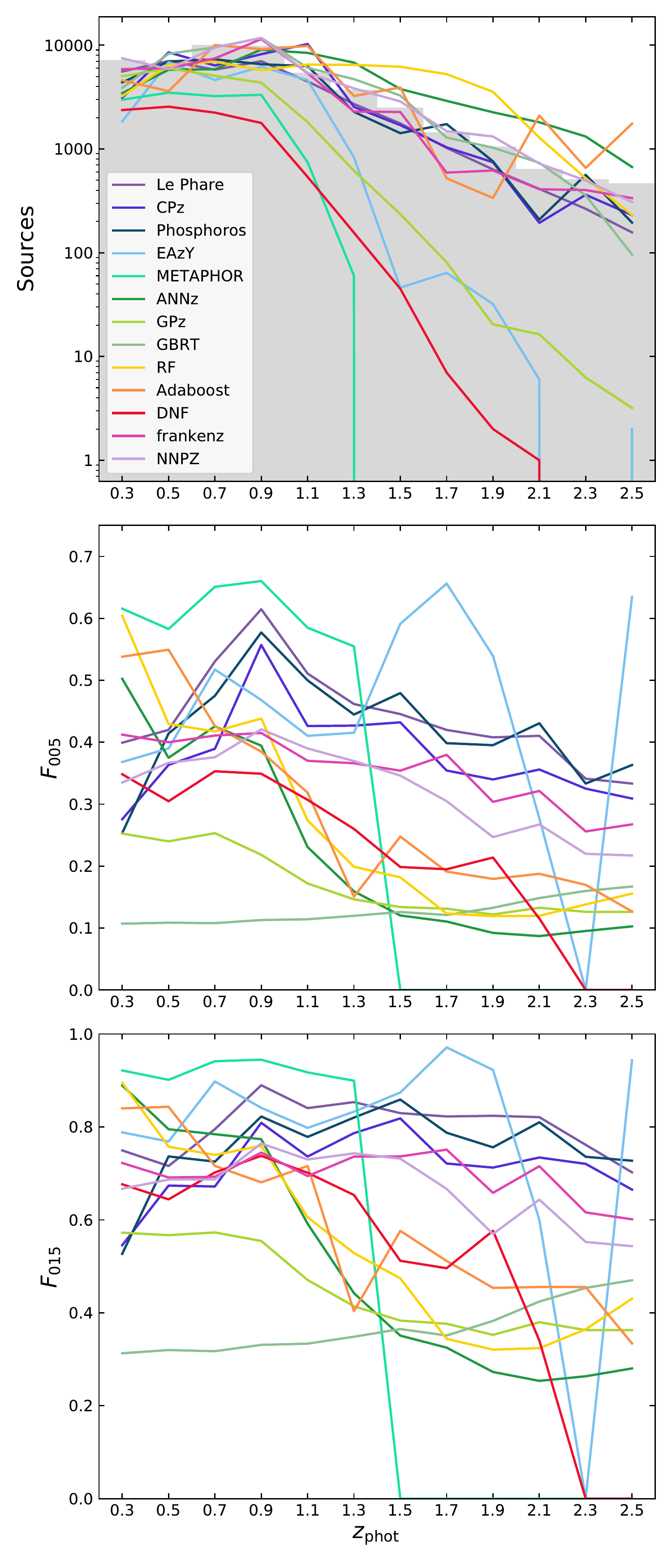}
    \caption{Same as Fig.~\ref{fig:specF005F015} for the L15 sample and the L15 photo-$z$s, instead of the spec-$z$s.}
    \label{fig:L15F005F015}
\end{figure}

\begin{figure*}
    \centering
    \includegraphics[width=\linewidth]{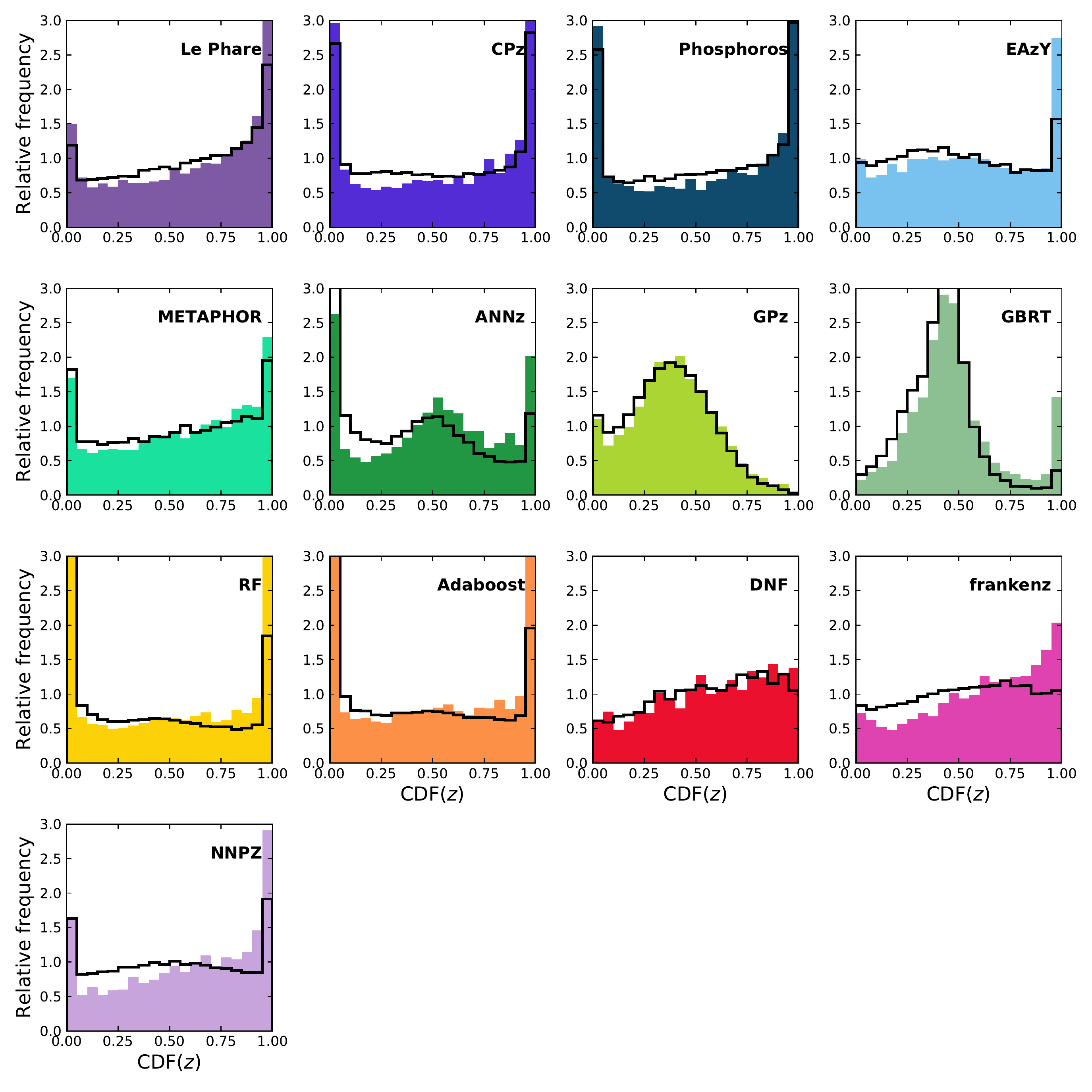} \\
    \caption{Probability integral transform (PIT) plots for all the methods, for the $\rm{USE}=1$ population and the \textit{Euclid} selection. Color histograms are the results for the weighted spectroscopic sample, while the black lines are the histograms for the L15 sample. }
    \label{fig:PITspec}
\end{figure*}

The metric we choose to assess the quality of the results is the one chosen to express the photo-$z$ requirements of \textit{Euclid}. The sources are first distributed in photo-$z$ bins depending on their point estimates. In each bin, the source PDZs, $P(z)$, are shifted by the values of the source spec-$z$s, $P(z-z_{\rm spec})$, in order to have the probability of the spec-$z$ at the origin. Then, all the shifted PDZs of each bin are stacked, using color-space weights for the spectroscopic sample (see Sect.~\ref{sec:results-pointEstimates}) and without weight for the L15 sample. For a bin centered on a redshift $z$, we compute the fractions of the stacked PDZ enclosed in $\pm0.05 (1+z)$ around its mode ($F_{005}$) and the one enclosed in $\pm0.15 (1+z)$ around the mode ($F_{015}$).
We note that integrating the stacked PDZs around the mode implies that it exists a method to correct their biases; the current baseline is to apply a calibration in color space using self-organizing maps \citep{Masters2015}. The quantities $F_{005}$ and $F_{015}$ measure the compactness of the distribution and can be compared to the scatter and outlier fraction of the point estimates. An $F_{005}$ larger than 68\% is the equivalent of the scatter being smaller than $0.05(1+z)$ when considering PDFs. Likewise an $F_{015}$ larger than 90\% corresponds to the fraction of outliers, which are objects with $|z_{\rm phot}-z_{\rm spec}|>0.15(1+z)$, being smaller than 10\%. Therefore $F_{005}$ and $F_{015}$ are the equivalent of the scatter and outlier fraction when dealing with PDZs, and the $F_{005} > 68\%$ and $F_{015}>90\%$ values are the equivalent to the requirements presented in the \textit{Euclid} Red Book \citep{Laureijs2011} when dealing with PDZs instead of point estimates.

Figure~\ref{fig:specF005F015} shows the weighted spectroscopic sample $F_{005}$ and $F_{015}$ fractions for all the tested methods in 12 photo-$z$ bins of width 0.2, from $z=0.2$ to $z=2.6$, as well as the number of sources per bin. In the distribution of sources per bin we see that the methods using strong rejection scheme, like \texttt{METAPHOR} or \texttt{EAzY}, provide very few, if any, predictions above $z=1$. 
The $F_{005}$ plot shows that template-fitting results and machine-learning results have distinct behaviors. \texttt{Phosphoros}, \texttt{Le Phare}, and \texttt{CPz} present a global level around $F_{005}\simeq0.4$ with a strong peak in $F_{005}$  at $0.6\lesssim z\lesssim 1.2$, and a small drop around $z=1.7$. The $F_{015}$ values of the template-fitting methods have roughly the same shape as the $F_{005}$ ones, with a base level of around $0.7$ and less pronounced peaks.
Some machine-learning methods (e.g., \texttt{GBRT}, \texttt{GPz}, and \texttt{DNF}) show $F_{005}<0.4$ everywhere, highlighting the difficulties of machine-learning algorithms in general in producing informative PDZs. However, other machine-learning methods (e.g., \texttt{ANNz}, \texttt{Adaboost}, \texttt{RF}, \texttt{METAPHOR}, \texttt{frankenz}, or \texttt{NNPZ}) produce good PDZs according to the $F_{005}$ and $F_{015}$ metrics, although they experience sharp drops of $F_{005}$ and $F_{015}$ above $z\simeq1.3$. We note that the machine-learning methods that show good results perform generally better than the template-fitting ones in the first three redshift bins, with the exception of \texttt{METAPHOR}, which shows better results than all the other methods until the $z=1.2$--$1.4$ bin, above which all sources are discarded. After that, the template-fitting methods show better results.
We notice the same behavior in the results for the L15 sample in Fig.~\ref{fig:L15F005F015}. In Fig.~\ref{fig:L15F005F015}, we see that the drop in $F_{005}$ around $z=1.7$ for the template-fitting results disappeared, possibly because the L15 PDZs are computed with similar template-fitting algorithms. We also see that the distinction between the template-fitting and machine-learning results is larger, due to a general decrease in $F_{005}$ for machine-learning approaches. The diminution of $F_{005}$ and $F_{015}$ for the L15 sample could be explained by the uncertainties of the L15 photo-$z$s, however template-fitting methods showing similar results in both weighted spectroscopic and L15 sample mitigates this possibility. The results on the L15 sample show the struggles of machine-learning methods to provide sensible results for a sample with a color space not matching the one they have been trained on.

The quality of the PDZs can be assessed using other metrics. We test the performance of the PDZ using probability integral transform plots (PIT plot, \citealt{Dawid1984,D'Isanto2018}). We compute the cumulative distribution function (CDF) at the true or L15 redshifts for all the sources $i$:
\begin{equation}
    C_{i}\equiv \mathrm{CDF}_{i}(z_{i}) = \int_{0}^{z_{i}} \mathrm{PDZ}_{i}(z)  \,{\rm d}z.
    \label{eq:CDF}
\end{equation}
 Figure~\ref{fig:PITspec} presents the histograms of $C_i$ for each of the tested methods, for both the color-space weighted spectroscopic sample and the L15 sample. If the PDZs correctly represent the probability distribution of the sources, the histograms should be flat. Outlier sources that have their spec-$z$s in the outskirts of their PDZs have their CDFs close to 0 or to 1 and produce the peaks at the edge of most of the histograms in Fig.~\ref{fig:PITspec}. U-shaped PIT plots, like those of \texttt{Phosphoros} or \texttt{CPz}, show that their PDZs are under-dispersed, meaning that they are in general too narrow. On the other hand, the PIT plots of \texttt{GPZ}, \texttt{ANNz}, or \texttt{GBRT} present a bump, indicating that the PDZs are over-dispersed, hence the PDZs are generally too broad. Biased PDZs produce PIT plots with a slope, like in the \texttt{Le Phare}, \texttt{METAPHOR}, \texttt{DNF}, or \texttt{Frankenz} histograms. We see that no method produces a perfectly flat PIT plot. We note also that there are no strong differences between the weighted spectroscopic sample and L15 sample PIT plots, with maybe the exceptions of \texttt{frankenz} and \texttt{NNPZ} that present flatter distributions for the L15 sample than on the weighted spectroscopic sample.

The other indicator often used to assess the quality of the PDZs is the continuous ranked probability score (CRPS, \citealt{Hersbach2000,D'Isanto2018}). It is defined as 
\begin{equation}
    \mathrm{CRPS}_{i} = \int_{-\infty}^{z_{i}} \mathrm{CDF}_{i}(z)^{2} \;{\rm d}z + \int_{z_{i}}^{+\infty} \left[\mathrm{CDF}_{i}(z)-1\right]^{2}  \,{\rm d}z,
\end{equation}
and should be close to zero for a narrow PDZ at the true redshift. However, the CRPS would increase both in the cases of PDZ at the wrong redshift or a broad PDZ around the true redshift. The median CRPS (weighted in the case of the spectroscopic sample) provided in Table~\ref{tab:CRPS} give an indication of the overall quality of the PDZs for each method. We use the median instead of the mean due to the high CRPS values for some few outliers (more than 30 times the mean value in some cases, see Fig.~\ref{fig:crps} for the complete distributions) increasing the mean value, which is thus less representative of the CRPS than the median for the majority of sources. Table~\ref{tab:CRPS} also reports the CRPSs obtained with precise (i.e., Dirac function) but biased PDZs or unbiased but dispersed PDZs (Gaussian with the true redshift as mean and a non-zero scatter), both tuned to provide values similar to those measured for the different methods. The CRPS is sensitive to both bias and scatter, with no possibility to distinguish between the two effects.
CRPS values in Table~\ref{tab:CRPS} show that for the spectroscopic sample the majority of methods present a median CRPS of around 0.08. Some methods provide better results like \texttt{METAPHOR}, \texttt{EAzY}, or \texttt{Le Phare} with median CRPS values of around 0.04 to 0.06, and some methods have larger mean CRPS like \texttt{ANNz}, \texttt{GPz}, \texttt{GBRT} or \texttt{RF}, with median CRPS above 0.1, meaning they provide less sensible PDZs which can be also deduced from $F_{005}$ and $F_{015}$ indicators for \texttt{GPz} and \texttt{GBRT}. 
For the L15 sample, there are no strong changes from the results on the weighted spectroscopic sample.

\begin{table}[h]   
    \centering
    \caption{Median continuous ranked probability score (CRPS) for the different algorithms using the \textit{Euclid} selection for the weighted spectroscopic and L15 samples.The last two rows present the CRPSs provided for sources in two cases: with infinitely precise Dirac PDZs but with a bias in range $0.06$--$0.18$ ; and with absolutely accurate ($\mathrm{bias}=0$) Gaussian PDZs with a scatter in range $0.15$--$0.7$.}
    \begin{tabular}{lcccc}
    \hline\hline
     & Spec. sample & L15 sample \\
    \hline
    \rule{0pt}{1.2em}\texttt{Le Phare} &  0.057 &  0.056 \\
    \texttt{CPz} & 0.087 & 0.091 \\
    \texttt{Phosphoros}  & 0.082 &  0.083 \\
    \texttt{EAzY} & 0.050 &  0.048 \\
    \texttt{METAPHOR} & 0.036 & 0.034 \\
    \texttt{ANNz} &  0.115 &  0.124 \\
    \texttt{GPz} & 0.116 &  0.113 \\
    \texttt{GBRT} &  0.166 &  0.165 \\
    \texttt{RF} & 0.107 &  0.119 \\
    \texttt{Adaboost} & 0.078 &  0.090 \\
    \texttt{DNF} &  0.076 &  0.072 \\
    \texttt{frankenz} &  0.071 &  0.072 \\
    \texttt{NNPZ}  & 0.084  & 0.081 \\
    bias ($0.06$--$0.18$)  & 0.040--0.160 & --- \\
    scatter ($0.15$--$0.7$) & 0.036--0.164  & ---\\
    \hline
    \end{tabular}
    \label{tab:CRPS}
\end{table}


\section{Discussion}
\label{sec:discussion}

We have performed extensive tests of the performance of 13 photo-$z$ algorithms using several metrics. One must keep in mind that all the analysis was done on the \textit{Euclid} shear sample, which only contains galaxies in a restricted photo-$z$ range of 0.2--2.6 (see Sect.~\ref{sec:data-shear}). This means that our results depend on the hypothesis that we are able to properly classify all the sources to obtain a pure sample of galaxies. If this is not the case, the resulting contamination would add an extra level of uncertainty in our results that is not captured by our tests.

\subsection{Point estimates}
\label{sec:discussion-PE}

The results on the full spectroscopic sample presented in Fig.~\ref{fig:PEstats} show that not all template-fitting methods provide similar results. Although \texttt{Le Phare}, \texttt{CPz}, and \texttt{Phosphoros} implement almost exactly the same algorithm, \texttt{Le Phare}'s results differ from these of the other codes, having slightly better results, with Fig.~\ref{fig:densityPlot} showing a strong similarity between \texttt{Phosphoros} and \texttt{CPz} outputs. The differences are therefore due to details of the configuration of the methods (i.e., data-driven, instead of code-driven differences) and we have checked that \texttt{Phosphoros} can reproduce almost exactly the \texttt{Le Phare} results if run under identical conditions. The first difference are the templates, \texttt{Le Phare} is including two additional templates (generated with an exponentially declining SFH) in addition to the 31 COSMOS template of \citet{Ilbert2009} that \texttt{Phophoros} and \texttt{CPz} use. A second difference is the point-estimate definition. Both \texttt{CPz} and \texttt{Phosphoros} use the PDZ mode, but \texttt{Le Phare} uses the PDZ median. Also, we note that \texttt{Le Phare} applied an absolute magnitude cut when running, added systematic errors to the magnitude errors, and applied a rejection for sources with overly broad PDZs.

For \texttt{CPz} and \texttt{Phosphoros}, having made roughly the same configuration choices, we see that they yield very similar results, as can be seen in Fig.~\ref{fig:densityPlot} and ~\ref{fig:PEstats}. The main difference between \texttt{Le Phare} or \texttt{CPz} and \texttt{Phosphoros} is the point estimate definition. This mostly impacts the point estimate metrics, but is less relevant in the rest of the analysis using PDZs (see Sect.~\ref{sec:discussion-PDZ}). Finally, the differences between \texttt{Le Phare} and \texttt{Phosphoros} show that there is some room for improvement in the configuration of the algorithms.

\texttt{EAzY} is a bit distinct from the other template-fitting codes: it uses a different set of templates than the 31 COSMOS templates, which it combines to fit the data; it uses a prior on the magnitudes in the $r$-band, which is not set by the other template-fitting codes; and it applies a different rejection scheme than the other codes, based on the odds of a PDZ being single-peaked. The results presented in Fig.~\ref{fig:densityPlot} are different than the results of the other template-fitting codes for these reasons. Despite these differences, it has similar performance to \texttt{Le Phare}, as seen in Fig.~\ref{fig:PEstats}.

We see in the top left panel of Fig.~\ref{fig:PEstats} that, for the whole sample, the lowest $\sigma$ and $\eta$ values are achieved by machine-learning algorithms (specifically \texttt{Adaboost} and \texttt{aNNz}). The rejection of the less reliable estimates can greatly improve the results for the point estimates. For example, \texttt{METAPHOR} sees its outlier fraction drop by about a factor of 6, and its scatter reduced by 25\% when applying a rejection based on its USE flag. Most rejection schemes seem to efficiently identify outliers, but have only a small effect on the scatter. However, the improvement in the outlier fraction comes at a price for completeness, since the most precise method after the rejection of flagged objects  (\texttt{METAPHOR}) discards 1/3 of all the sources and the second one (\texttt{DNF}) rejects half of them. Figure~\ref{fig:PEstats} shows also that the \textit{Euclid} selection leads to an improvement of the outlier fraction, but mainly for the methods that do not make any rejection of possibly wrong results on their own. This indicates that a large fraction of the outliers are located in redshift ranges outside of the \textit{Euclid} cosmic-shear target region.

When weighting the results using the \citet{Lima2008} weighting scheme or using the L15 sample, we see that most of the machine-learning results degrade greatly if the methods do not apply strong rejection. Training is indeed very poor in areas of color space that have a large weight. This can be mitigated if algorithms are able to identify and reject objects in these areas, which is especially the case for \texttt{METAPHOR}. On the other hand, \texttt{Adaboost}, \texttt{RF}, and \texttt{ANNz} are strongly penalized by the absence of rejection in their configurations. Another mitigation measure is the use of L15 photo-$z$s in the training, as is the case for \texttt{GBRT} and \texttt{NNPZ}. For these examples, the results on the L15 sample are even better than those on the color-space weighted sample.

Color-space weight and L15 scores are mostly similar, except for \texttt{Adaboost}, \texttt{RF}, and \texttt{ANNz}, which shows that these methods are able to make reasonable predictions even with few training objects, but that there are significant areas of the color space without any spec-$z$s. This could mean that the color-space weights are overestimating the results that the methods would have on the full photometric sample. An alternative explanation is that this could be due to bad L15 photo-$z$s, since the comparison between L15 photo-$z$s and spec-$z$s in the validation catalog shows a scatter $\sigma=0.013$ and an outlier fraction $\eta=11.0\%$, this explanation cannot be excluded. However the consistency of the L15 and color-space results for most algorithms show that these errors, if they are significant, happen essentially where spec-$z$ coverage is scarce. Template-fitting results do not seem to suffer as strongly as machine-learning results when applied to the color-space weighted or the L15 samples. Template-fitting appears to be able to provide sensible results even in the areas of the color space not covered by spectroscopic-redshifts, but we point out that template-fitting methods might perform well in these areas of color space because L15 photo-$z$s used the same algorithm and the same templates as \texttt{Le Phare}, \texttt{Phosphoros}, and \texttt{CPz}. However, differences in depth and wavelength coverage somewhat mitigate this issue.

\subsection{PDZs}
\label{sec:discussion-PDZ}

Although PIT plots and CRPSs have been used in recent works as PDZ quality indicators \citep[e.g.,][]{Tanaka2018,Pasquet2019}, they are not very useful indicators of the precision of the PDZs. CRPS is sensitive to both bias and scatter, in a way that makes the two effects difficult to disentangle. PIT checks that the individual spec-$z$s can be on average drawn from the PDZs, but it does not say anything about the quality of the predictions. \citet{Schmidt2020} give the example of a method without any predictive power, but with a perfect PIT, meaning that a method providing a perfect PIT plot can lead to a bad FoM. The same behavior can be expected from the CRPS, since the same CRPS values dominated either by the bias or the precision of the PDZs will provide different FoMs.
Nevertheless the general shapes of the PIT plots give some indications of whether the PDZs are over- or under-dispersed, biased, or outliers. Most PIT plots in Fig.~\ref{fig:PITspec} appear reasonable, with the exception of  \texttt{GPz}, \texttt{GBRT}, and \texttt{ANNz}. One the other hand, none of the PIT histograms are flat. Some methods still manage to provide fairly flat but biased PIT plots, especially for the L15 sample (like \texttt{frankenz}), or are slightly concave in the center and with small peaks toward the edges (like \texttt{NNPZ} or \texttt{EAzY}). This means that no method can produce PDZs that are in total agreement with the spec-$z$ or the L15 photo-$z$ distributions. This is not a fatal issue, however, since there are ways to correct the PDZs in order to flatten the PIT plot \citep{Bordoloi2010,Gomes2018} and to correct for most of the bias. In addition, the \textit{Euclid} science goals do not require the determination of the true $n(z)$, but only of the average redshifts in the tomographic bins, which is a significantly less ambitious goal that can be reached, for example, using self-organizing maps as proposed by \citet{Masters2015}. 

In the context of the \textit{Euclid} mission, we define new estimators of the photo-$z$ precision that are insensitive to the bias. The \textit{Euclid} requirements are expressed using the $F_{005}$ and $F_{015}$ definitions, which consider the PDZs around the mode of the distributions, making these metrics sensitive only to the precision (the width) of the PDZs. They can also be easily associated with the scatter and outlier fraction of point estimates. For these reasons, we focus on the $F_{005}$ and $F_{015}$ measurements for the different methods. The binning of the results in tomographic bins presented in Fig.~\ref{fig:specF005F015} and~\ref{fig:L15F005F015} allows us to see more clearly what was hinted in Fig.~\ref{fig:densityPlot}, namely that strongly-rejective  methods are mainly rejecting sources with redshifts $z>1$. This is the case for \texttt{METAPHOR}, which does not provide any results above the bin at $z=1.2$--$1.4$, neither for the weighted spectroscopic sample, nor the L15 sample. However, this strong rejectivity results in high scores for the metrics in the domain in which results are provided. Figures~\ref{fig:specF005F015} and~\ref{fig:L15F005F015} show that the methods that have poor results in their PIT plots and CRPSs do not perform well on the $F_{005}$ and $F_{015}$ metrics (e.g., \texttt{GBRT}, \texttt{GPz}, or \texttt{ANNz}). Figure~\ref{fig:specF005F015} shows that machine-learning methods tend to perform better than template-fitting ones in the redshift range of $z<0.8$, but perform worse above this redshift. Using the L15 sample (Fig.~\ref{fig:L15F005F015}), the gap between the results of machine-learning approaches and those of template-fitting is larger than that obtained from the spectroscopic sample. This indicates that (perhaps unsurprisingly) the machine-learning algorithms also have more difficulty in providing sensible PDZs for sources that are rarely or not at all represented in the training sample. An increase in the redshift coverage of the color space is needed to more properly train the machine-learning methods. 
Ongoing and future spectroscopic survey programs (e.g., C3R2, \citealt{Masters2017,Masters2019,Guglielmo2020}) will increase the color space coverage  with high-quality spectroscopic redshifts, thus the performance of machine-learning algorithms is expected to improve over time due to a better training sample. However, it is not clear that the number of spec-$z$s will be sufficient to both train the machine-learning methods and calibrate the bias of the photo-$z$s without introducing new sources of bias. 

Template-fitting codes use an explicit model of the galaxy SEDs, and thus they provide better results at high redshift than machine-learning algorithms, which rely on training sources in this regime. However, at redshifts $z<0.5$, all the template-fitting methods are outmatched by machine-learning methods. The superior results of machine-learning approaches at low redshifts show that the photometry does contain enough information to constrain the photo-$z$s. Nevertheless template-fitting methods have trouble in this region. This may result from a lack of valid templates at low redshift, or it may be due to a lack of proper priors, which are present in machine-learning methods in an implicit way due to the training data set containing mostly sources with low redshifts.

\begin{figure}
    \centering
    \includegraphics[width=\linewidth]{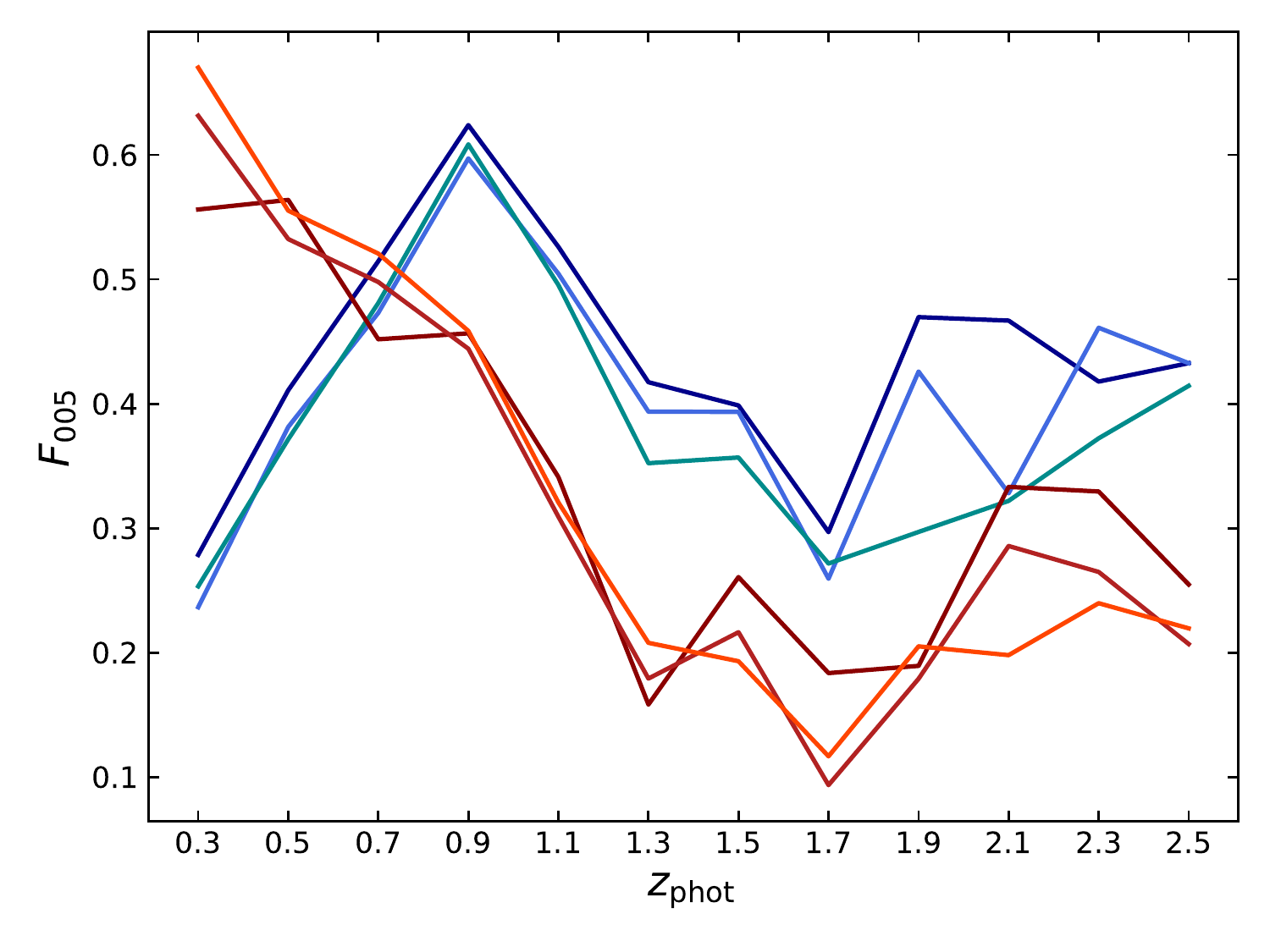}
    \caption{$F_{005}$ plot on the weighted spectroscopic sample showing the impact of the definition of the point estimates used to sort the sources into the redshift bins for \texttt{Phosphoros} and \texttt{Adaboost}.}
    \label{fig:modeormedian}
\end{figure}

\begin{figure*}
    \centering
    \includegraphics[width=\linewidth]{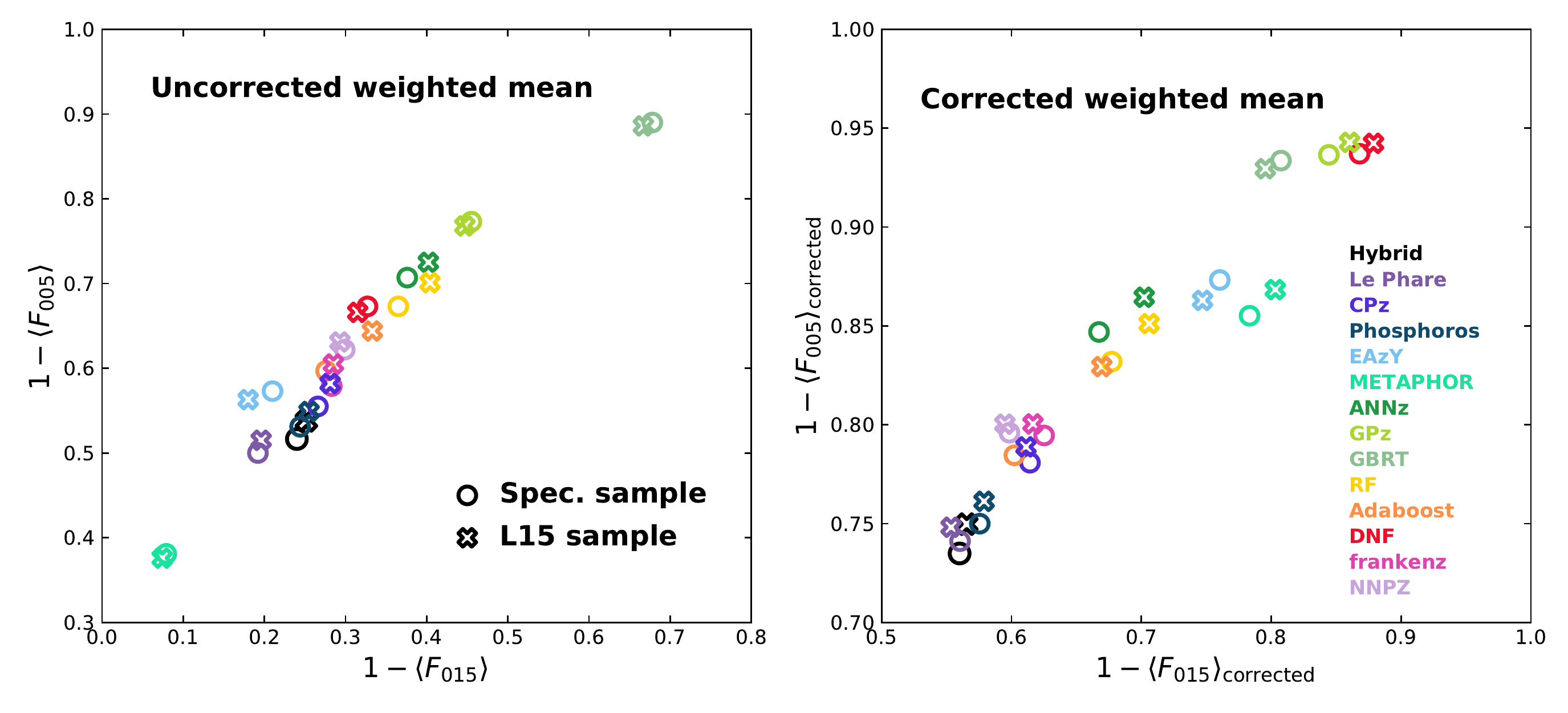}
    \caption{PDZs metrics summarized by averaging the $F_{005}$ and $F_{015}$ values on all the bins with different weighting schemes. The axes are $ 1-\langle F_{005}\rangle$ and $1-\langle F_{015}\rangle$ to mimic the usual $\sigma$--$\eta$ plots in Fig.~\ref{fig:PEstats}. \textit{Left}: Results per bin weighted by the fraction of sources in the bin compared to the total number of sources kept by each methods (see Eq.~\ref{eq:score1}). \textit{Right}: Results of all the methods when correcting the $\langle F_{005}\rangle$ and $\langle F_{015}\rangle$ of each bin by the square root of the ratio of good sources in the bins to the number of sources that truly belong to the bin (see Eq.~\ref{eq:score3}). In each plot we include the results for the hybrid method (in black, see Sect~\ref{sec:discussion-hybrid}) for the weighted spectroscopic sample on the L15 sample. }
    \label{fig:metric}
\end{figure*}

In Sect.~\ref{sec:discussion-PE}, we explained that different definitions of point estimates can lead to differences in results. Our PDZ metrics are still sensitive to the point estimate variations, since we use them to sort the sources within the tomographic bins. Figure~\ref{fig:modeormedian} shows an example of the impact of the definition of the point estimates on the $F_{005}$ fraction for \texttt{Phosphoros} and \texttt{Adaboost}. We observe some differences between the results, mostly at redshifts $z>1$. In that range of redshift, the mode seems to be the point-estimate that provides the best results. For the $z<1$ redshift range, we see very little variation of the results with the definition of the point estimates.

\subsection{Maximizing the proper metric}
\label{sec:discussion-metrics}

In the context of \textit{Euclid}, the metric that is maximized is the dark-energy figure of merit (see \citealt{Laureijs2011} for a detailed description). The dark energy FoM increases with the quality of the weak-lensing signal, and this signal depends on the quality of the photo-$z$s, but also on the number of sources for which the photo-$z$s are measured.\footnote{it clearly also depends on other parameters, but we focus here on the effects on which the photo-$z$ algorithms have influence.} The requirement presented in the \textit{Euclid} Red Book is that the galaxy density must be over $30$\,galaxies per arcmin$^2$.

The results presented in Sect.~\ref{sec:results-pointEstimates}~and~\ref{sec:results-PDFs} show that a rejection of the sources on which to carry out the analysis allows the methods to improve the precision of the redshifts. However, the $F_{005}$ and $F_{015}$ metrics are not sensitive to the loss of information resulting from this rejection. The same problem is true for PIT and CRPS. Figures~\ref{fig:specF005F015} and~\ref{fig:L15F005F015} show that some methods, such as \texttt{METAPHOR}, leave some tomographic bins completely  unpopulated. This means that no weak-lensing analysis can be performed at these redshifts, resulting in a strong loss of FoM and a failure to meet the \textit{Euclid} mission requirements if such drastic rejection is made.

We use two methods of averaging the $F_{005}$ and $F_{015}$ metrics over the tomographic bins (Fig.~\ref{fig:metric}). First, the weight applied to $F_{005}$ and $F_{015}$ in each bin is the number of objects put in this bin by a given photo-$z$ method, that is,
\begin{equation}
    \langle F_{0XX}\rangle = \frac{1}{N_{\mathrm{USE=1}}} \sum^{\rm{bins}}_{i}{ F_{0XX,i} N_{\rm sources \textit{,i}}},
    \label{eq:score1}
\end{equation}
where $F_{0XX,i}$ is either $F_{005}$ or $F_{015}$ (or any other desired value) in a bin $i$, $N_{\rm{sources}\textit{,i}}$ is the number of sources in a bin and $N_{\rm{USE}=1}$ is the total number of sources in all the bins. These weights roughly reproduce the standard estimators for point estimates $\sigma$ and $\eta$ in the case of PDZs, since they are averaged over all objects. The $\langle F_{0XX}\rangle$ metric does not penalize methods with strong rejection because the empty bins have null weight in the average computation, thus $\langle F_{0XX}\rangle$ does not reflect the negative impact that underpopulated, or even empty, tomographic bins can have on the weak-lensing analysis. Using this average, the best methods seem to be \texttt{METAPHOR}, \texttt{Le Phare}, and \texttt{Phosphoros}.

Another way to produce an average $\langle F_{0XX}\rangle$ would be to assume that each tomographic bin has the same weight in the weak-lensing signal, which translates into unweighted averages of $F_{005}$ and $F_{015}$. This would give a penalty to methods rejecting all objects in a given bin or to methods that are particularly poor in some redshift range (typically machine-learning at high $z$). However, it would does not impact results with underpopulated bins that could obtain good $F_{005}$ and $F_{015}$ values, but not enough sources to improve the weak-lensing analysis results. For this reason, the metric must take into account the population of the tomographic bins. To do so, a correction is introduced that depends on the fraction of objects correctly assigned to the bin:
\begin{equation}
    \langle F_{0XX}\rangle_{\mathrm{corrected}} =  \frac{1}{N_{\mathrm{bins}}}\sum^{\mathrm{bins}}_{i}{ F_{0XX,i} \sqrt{\frac{N_{\mathrm{good},i}}{N_{\mathrm{true},i}}}},
    \label{eq:score3}
\end{equation}
where $N_{\rm{good} \textit{,i}}$ is the number of sources that have been correctly placed in the bins $i$, and $N_{\rm{true} \textit{,i}}$ is the true number of sources in bin $i$ (see Fig.~\ref{fig:goodfraction} for the values of the fractions per bin). The square root is applied to reproduce the dependency of the increase in precision with the number of objects. Using the fraction of ``good'' sources compared to the number of ``true'' sources in the bin penalizes underpopulated but not empty bins with high fraction values. It also ignores outliers falling in the bins, which could artificially boost the scores of the bins. Figure~\ref{fig:metric} (right panel) shows the result of this correction. Template-fitting methods (\texttt{Le Phare} and \texttt{Phosphoros}) present the best results, but some machine-learning methods being less penalized, such as \texttt{Adaboost} and \texttt{NNPZ}, also yields good performance. Nevertheless, this correction is a simple and intuitive way to estimate the trade-off between the precision of the photo-$z$s and the number of sources considered, but the proper metrics to consider here would take into account the weight of each sources and tomographic bins in the estimation of the weak-lensing signal.

It would be desirable to apply a penalty similar to that used in Eq.~(\ref{eq:score3}) for the PIT and CRPS metrics. Unfortunately, there is no sensible way to estimate how the loss of sources would affect them, and neither the CRPS, nor any statistics derived from the PIT can be unambiguously translated into a FoM.

\subsection{Improving the results}
\label{sec:discussion-hybrid}

\begin{figure}
    \centering
    \includegraphics[width=\linewidth]{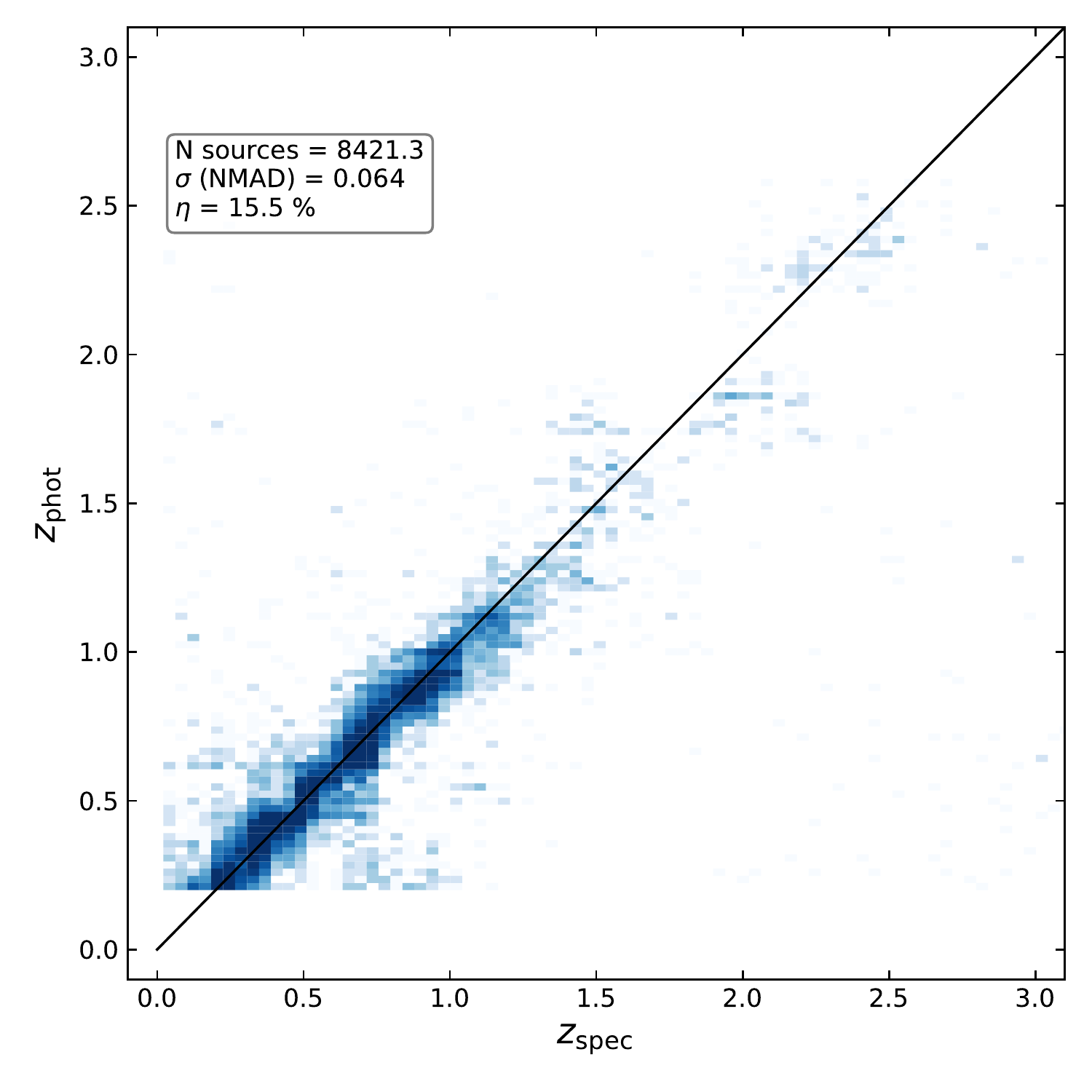} \\
    \includegraphics[width=\linewidth]{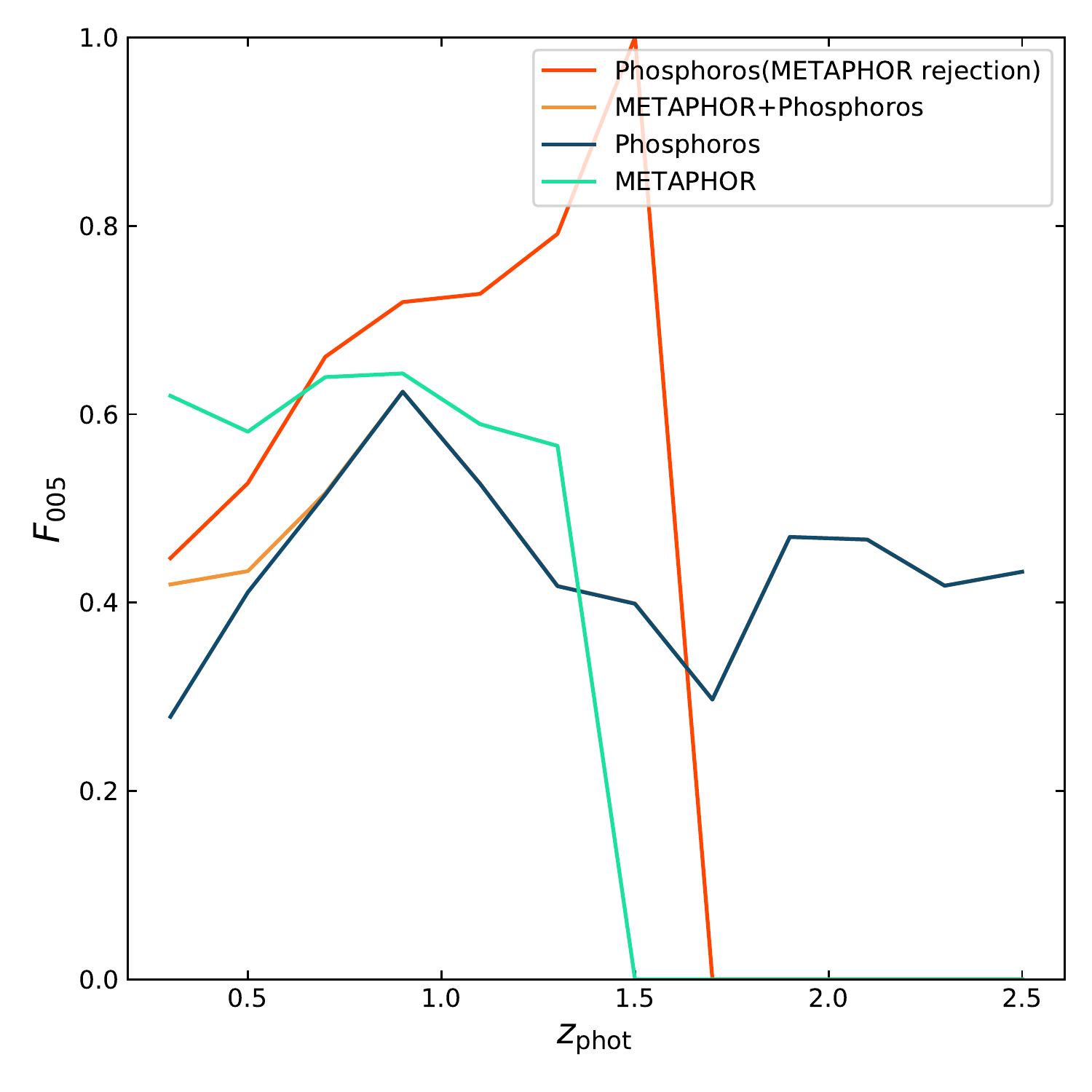}
    \caption{\textit{Top}: scatter plot of results of the combination of \texttt{METAPHOR} and \texttt{Phosphoros} point estimates on the weighed spectroscopic sample and the \textit{Euclid} selection. \textit{Bottom}: $F_{005}$ for \texttt{METAPHOR} and \texttt{Phosphoros} results along with the combination of their results. The red curve is the results of \texttt{Phosphoros} when the \texttt{METAPHOR} rejection scheme is applied to it. The orange curve shows the combination of the results of \texttt{METAPHOR} and \texttt{Phosphoros} in the first two bins, then the curves merges with the typical \texttt{Phosphoros} one. }
    \label{fig:metaphosphoros}
\end{figure}

Each methods has its advantages and disadvantages, and thus performs efficiently in different regimes. Machine-learning methods are based on a training sample and their results depend strongly on the quality of this training. Template-fitting methods do not have this problem and perform relatively well for sources in regions of the color space with a sparse redshift coverage. However, Fig.~\ref{fig:specF005F015} shows that they can be outmatched by machine-learning in the well covered regions of the color space. As mentioned earlier, both types of method can be improved separately (see Sect.~\ref{sec:discussion-PDZ}). However, Fig.~\ref{fig:specF005F015} also shows that some methods (such as \texttt{METAPHOR}) are able to substantially improve  the precision of their results by accurately predicting when a result is a probable outlier. In Sect.~\ref{sec:discussion-metrics} we see that non-rejective template-fitting methods (such as \texttt{Phosphoros} or \texttt{Le Phare}) are performing well with a metric approximating the effect of photo-$z$s on the weak-lensing analysis, whereas very precise but highly rejecting methods (such as \texttt{Metaphor} or \texttt{DNF}) not providing results above redshift 2 are incompatible with the goals of the weak-lensing analysis. 
Nevertheless, the ability of \texttt{METAPHOR} to predict outliers could be used to improve the results. For that, we first must check if \texttt{METAPHOR} really surpasses other methods for the objects for which it provides results. Figure~\ref{fig:metaphosphoros} presents the $F_{005}$ curve for \texttt{Phosphoros} restricted to sources that have a USE flag 1 with \texttt{METAPHOR}. We see that the  \texttt{Phosphoros} results are greatly improved up to the point where the rejection discards all the sources, above $z=1.5$. Also, we note that the results of \texttt{METAPHOR} are only better than the results of \texttt{Phosphoros} with the same rejection in the first two bins of the \textit{Euclid} redshift range. Thus, to improve results, we propose a hybrid photo-$z$ algorithm as follow: we use the results of \texttt{METAPHOR} when its USE flag is 1 and its predicted photo-$z$ is below $z=0.6$, otherwise we use the \texttt{Phosphoros} results if not.

The results of this hybrid method are presented in Fig.~\ref{fig:metaphosphoros} with a $F_{005}$ plot and a scatter plots. The scatter plot in Fig~\ref{fig:metaphosphoros} shows that the hybrid approach has similar scatter, but a slightly smaller outlier fraction than \texttt{Phosphoros} (see Table~\ref{tab:statsWeight}), which results in an increase in the number of sources in the Euclid sample at low redshift. The results of the hybrid method are the best ones of all methods that do not reject any source. The resulting $F_{005}$ curve in Fig~\ref{fig:metaphosphoros} differs from the \texttt{Phosphoros} one only in the two first bins, since above $z=0.6$  only \texttt{Phosphoros} results are used. The curve still remain under the \texttt{METAPHOR} one, but a solution is provided for all the sources in the sample. We can see the improvement it brings in Fig~\ref{fig:metric}, that shows the results of this approach in the averaged $F_{0XX}$ metrics.
The corrected mean $F_{005}$ of this hybridization is $\langle F_{005}\rangle_{\rm{corrected}}=0.27$ on the weighted spectroscopic sample. This result is better than all the results presented in Fig.~\ref{fig:metric}. The increase compared to \texttt{Phosphoros} is only due to the improvement in $F_{005}$ in the first two bins. 

This method of combining the results of machine-learning and template-fitting seems promising (e.g., \citealt{Brodwin2006,Duncan2018}) and should be explored further, possibly with a better criterion for selecting the predictions from the machine-learning or the template-fitting algorithms.


\section{Summary and conclusions}
\label{sec:conclusion}

Thirteen different photo-$z$ methods, both template-fitting and machine-learning based, have been tested on \textit{Euclid}-like data. Each method has provided each source a point estimate redshift, a PDZ, and a USE flag, allowing them to reject sources considered problematic. Their results have been compared through different metrics with the aim of assessing the impact of the provided photo-$z$s on the \textit{Euclid} cosmic-shear analysis. For this reason, we analyze the results for galaxies in the 0.2--2.6 photo-$z$ range only. The tests we have conducted here have therefore little relevance for the study of high-redshift galaxies, for instance. We have further assumed that a proper classification between stars, galaxies and AGNs has already been done, and that the photo-$z$s can be calibrated independently of the photo-$z$ algorithm.

The results show that adopting stringent rejection criteria can be very efficient in reducing the outlier fraction. Some methods are quite successful in accurately identifying sources with reliable photo-$z$s. However, the drawback of such rejection is a loss of completeness for further analysis, which can be incompatible with some science goals, in particular weak-lensing tomography.   

To assess the quality of PDZs, the PIT plot and CRPS are standard metrics. They must be considered together, since PIT only indicates whether the true redshifts are collectively compatible with being drawn from the PDZs, and CRPS is sensitive to both the bias and scatter of the results. However, its sensitivity to the bias, which cannot be disentangled from the effect of the scatter, is not suited for our analysis that only focuses on the precision of the results. This leads us to define the fraction metrics ($F_{005}$ and $F_015$) related to the \textit{Euclid} requirements on the precision of the PDZs. The fraction metrics can also be corrected to take in account the loss in completeness that is due to rejection schemes of the different methods.

Analysis of the PDZ results shows that producing sensible PDZs is not straightforward for machine-learning methods, as several of them do not manage to provide good PDZs, regardless of the indicator used to assess their quality. Machine-learning methods also struggle to make good predictions over large areas of the color space, in particular for $z>1$ or regions scarcely covered by spectroscopic information, even though the COSMOS training sample is one of the most complete spec-$z$ samples currently available.

However, in regions of color space well covered by spec-$z$s, machine-learning methods (e.g., \texttt{METAPHOR} or \texttt{Adaboost}) seem to perform the best. With an appropriate spec-$z$ sample, they could outmatch all the other methods. However, the construction of a perfect training sample covering the full color space at the limiting depth of the surveys with sufficiently numerous spec-$z$, remains intractable. Using L15 photo-$z$s for this purpose is a possible compromise, as shown by \texttt{NNPZ} in particular.

Template-fitting methods show more consistent results than machine-learning over the full photometric sample; however, they seem unable to use the full information contained in the photometry at low redshifts. The reason for this behavior must be understood whether it is a lack of templates or a better definition of priors in order to improve these methods.

Taking into account the properties of the output photo-$z$s, the driver of the choice of algorithm is the use made of them. The metrics used to compare the results of the algorithms depend on the purpose of the photo-$z$s and must reflect the impacts they will have on the science case foreseen. For weak-lensing studies, completeness is needed, and template-fitting appears to perform best when assessing both the precision of the photo-$z$s and their numbers. However, if high precision and purity are required, then machine-learning seems better in those aspects, especially when they implement rejection of poor predictions.

Thanks to the capability of rejecting probable outliers, we can overcome the limits of both approaches and combine the high precision of machine-learning and the completeness of template-fitting. This combination of results shows better average photo-$z$ precision than any method alone, while preserving the completeness of the considered sample of galaxies, hence solving the issue of the loss of sources, which impacts negatively the weak-lensing analysis and the \textit{Euclid} dark energy FoM. Further work is required to determine the optimal combination between template-fitting and machine-learning algorithms.

\begin{acknowledgements}
      GD thanks Douglas Scott for his very helpful comments on the manuscript.
      GD and AG acknowledge the support from the Sinergia program of the Swiss National Science Foundation.
      Part of this work was supported by the German
      \emph{Deut\-sche For\-schungs\-ge\-mein\-schaft, DFG\/} project
      number Ts~17/2--1.
      MB acknowledges the financial contribution from the agreement \textit{ASI/INAF 2018-23-HH.0, Euclid ESA mission - Phase D} and the \textit{INAF PRIN-SKA 2017 program 1.05.01.88.04}. SC acknowledges the financial contribution from FFABR 2017.
      \AckEC
\end{acknowledgements}

%
%

\bibliographystyle{aa} 
\bibliography{references} 

\begin{appendix}

\section{The $VIS$ simulation software}
\label{sec:visSoft}

The $VIS$ simulation software takes as input a high-resolution image (using the \textit{HST} ACS F814W in the COSMOS field in this specific case) and manipulates it in order to obtain a simulated image with the desired features (i.e., degrading the resolution to the expected $VIS$ resolution and adding noise). The pipeline implements four processing steps executed in the following sequence:
\begin{itemize}
\item Mkkernel, generates an analytical (Gaussian) kernel according to the input image PSF and the PSF requested for the simulated one;
\item Convolve, operates the convolution from the input image to the convolved one according to the previously generated kernel;
\item Swarp, performs the rebinning of the convolved image to the required pixel scale;
\item Mknoise, Gaussian noise is added in each pixel to reproduce the desired depth in the output.
\end{itemize}

The original ACS F814W image has a non-uniform depth, and particular care has been devoted to the noise addition: Gaussian noise is added to each pixel according to a scaling factor that takes into account the pixel-to-pixel variation of the original image depth. The resulting rms map is an image with constant value in the portion covered by the observation and has a constant value of $10^{16}$ outside. The rms map value is the result of the following equation:
\begin{equation}
{\rm rms}_{\rm out} = \frac{10^{0.4\,({\rm ZP} -m_n)}}{{\rm S/N} \, \sqrt{\pi} \, \frac{n{\rm FWHM}}{2{\rm pxs}}} \, ,
\end{equation}
where $m_n$ is the reference magnitude (at the given S/N) measured in $n$ (1, 2, or 3) times the PSF FWHM. ZP and pxs are the zeropoint and the pixelscale of the image, respectively. Where the original image has been found to be shallower than requested no Gaussian noise has been added and the rms value has not been modified. Currently, photon noise from the sources is not included.

\section{Point estimate metric tables}
\label{sec:pointEstimateTables}

In Sect.~\ref{sec:results-pointEstimates}, we present the results of the different methods in several conditions, using multiple selections (e.g., USE flag or \textit{Euclid} selection) and comparing the photo-z's to different reference redshifts. These results are summarized in Fig.~\ref{fig:PEstats}, using the values compiled in Tables~\ref{tab:statsAll},~\ref{tab:statsEuclid},~\ref{tab:statsWeight}, and~\ref{tab:stats30bands} in this section.

Table~\ref{tab:statsAll} contains the scatter ($\sigma_{\rm all}$) and outlier fraction ($\eta_{\rm all}$) for all the methods, considering the complete spectroscopic shear sample (12463 sources) present in the validation catalog. This table also shows for each method the number of sources remaining after the USE flag selection is applied ($N_{\rm{USE=1}}$) and the $\sigma_{\rm{USE=1}}$ and $\eta_{\rm{USE=1}}$ associated with this selection.

\begin{table}[h]
    \centering
    \caption{Point estimate statistics for the spectroscopic sample. The scatter ($\sigma$) and the outlier fraction ($\eta$) are given in the case of no rejection with subscript ``all'' (i.e., 12\,463 sources), and in the case of rejection with the USE flag with subscript ``USE=1''. In the second case, the number of selected source is also displayed ($N_{\rm{USE}=1}$). }
    \begin{tabular}{lccccc}
    \hline\hline
         & $\sigma_{\rm{all}}$ & $\eta_{\rm{all}}$ & $N_{\rm{USE}=1}$ & $\sigma_{\rm{USE}=1}$ & $\eta_{\rm{USE}=1}$ \\
         &  & [\%] & & & [\%] \\
    \hline
        \rule{0pt}{1.2em}\texttt{Le Phare} & 0.046 & 12.0 & 11377 & 0.043 & \phantom{0}8.1\\
        \texttt{CPz} & 0.066 & 15.5 & 10841 & 0.066 & 15.6 \\
        \texttt{Phosphoros}   & 0.066 & 15.8 & 12463 & 0.066 & 15.8 \\
        \texttt{EAzY} & 0.058 & 15.4 & \phantom{0}9594 & 0.047 & \phantom{0}6.2 \\
        \texttt{METAPHOR} & 0.051 & 16.6 & \phantom{0}8302 & 0.037 & \phantom{0}2.8 \\
        \texttt{ANNz} & 0.048 & 10.0 & 12463 & 0.048 & 10.0 \\
        \texttt{GPz} & 0.078 & 14.2 & 10676 & 0.069 & \phantom{0}9.3 \\
        \texttt{GBRT} & 0.058 & \phantom{0}9.7 & 12311 & 0.058 & \phantom{0}9.2 \\
        \texttt{RF} & 0.052 & 11.5 & 12463 & 0.052 & 11.5 \\
        \texttt{Adaboost} & 0.046 & \phantom{0}9.2 & 12463 & 0.046 & \phantom{0}9.2  \\
        \texttt{DNF} & 0.055 & 12.2 & \phantom{0}5520 & 0.041 & \phantom{0}5.9 \\
        \texttt{frankenz} & 0.068 & 28.3 & \phantom{0}9661 & 0.042 & \phantom{0}8.8 \\
        \texttt{NNPZ} & 0.061 & 12.1 & 12463 & 0.061 & 12.1\\
    \hline
    \end{tabular}
    \label{tab:statsAll}
\end{table}

The results for the \textit{Euclid} selection (i.e., being part of shear sample, with photo-$z$ in the range 0.2--2.6, and $\rm{USE}=1$) are listed in Table~\ref{tab:statsEuclid}. The column $N_{\rm{Euclid}}$ shows the number of sources for each method. The scatter  $\sigma_{\rm{Euclid}}$ and the outlier fraction $\eta_{\rm{Euclid}}$ are also reported.

\begin{table}
\centering
\caption{Point estimate statistics for the \textit{Euclid} sample. Number of sources $N_{\rm{Euclid}}$, scatter $\sigma_{\rm{Euclid}}$, and outlier fractions $\eta_{\rm{Euclid}}$ are provided. }
\begin{tabular}{lcccccc}
\hline\hline
 & $N_{\rm{Euclid}}$ & $\sigma_{\rm{Euclid}}$ & $\eta_{\rm{Euclid}}$ \\
 & & & [\%]& & \\ 
\hline
\rule{0pt}{1.2em}\texttt{Le Phare}  & 10607 & 0.041 & \phantom{0}6.9 \\
\texttt{CPz} & \phantom{0}8985 & 0.055 & \phantom{0}9.7  \\
\texttt{Phosphoros}  & 10140 & 0.055 & \phantom{0}8.7 \\
\texttt{EAzY} & \phantom{0}9286 & 0.046 & \phantom{0}6.2 \\
\texttt{METAPHOR} &  \phantom{0}7865 & 0.036 & \phantom{0}2.7  \\
\texttt{ANNz} & 12012 & 0.048 & 10.1  \\
\texttt{GPz} & 10208 & 0.068 & \phantom{0}9.2  \\
\texttt{GBRT} & 11978 & 0.057 & \phantom{0}9.3  \\
\texttt{RF} & 11955 & 0.05\phantom{0} & 10.5   \\
\texttt{Adaboost} & 12008 & 0.045 & \phantom{0}9.0  \\
\texttt{DNF} & \phantom{0}5101 & 0.042 & \phantom{0}5.8  \\
\texttt{frankenz} & \phantom{0}8870 & 0.041 & \phantom{0}8.1  \\
\texttt{NNPZ} & 11501 & 0.059 & 11.1 \\
\hline
\end{tabular}
\label{tab:statsEuclid}
\end{table}

Table~\ref{tab:statsWeight} presents the results for the \textit{Euclid} selection after re-weighting it to be more representative of the photometric sample, using the \cite{Lima2008} scheme. For each method we provide $N_{\rm{color-space}}$, which is the sum of the computed weights of all the selected sources, along with the weighted scatter $\sigma_{\rm{color-space}}$ and outlier fraction $\eta_{\rm{color-space}}$.

\begin{table}[h]
    \centering    
    \caption{Point estimate statistics for the color-space weighted-sample. The sum of the weights of sources in each method selection $N_{\rm{color-space}}$, scatter $\sigma_{\rm{color-space}}$, and outlier fractions $\eta_{\rm{color-space}}$ are provided.}
    \begin{tabular}{lccccccccccccc}
    \hline\hline
     & $N_{\rm{color-space}}$ & $\sigma_{\rm{color-space}}$ & $\eta_{\rm{color-space}}$  \\
    &  & & [\%]&  \\ 
    \hline
    \rule{0pt}{1.2em}\texttt{Le Phare} & \phantom{0}7644.0 & 0.056 & 13.4  \\
    \texttt{CPz} & \phantom{0}8569.7 & 0.077 & 21.1  \\
    \texttt{Phosphoros} & \phantom{0}8084.1 & 0.067 & 17.1 \\
    \texttt{EAzY} & \phantom{0}5772.5 & 0.062 & 13.2   \\
    \texttt{METAPHOR} & \phantom{0}3039.6 & 0.04\phantom{0} & \phantom{0}3.1   \\
    \texttt{ANNz} & 10564.7 & 0.091 & 26.1  \\
    \texttt{GPz} & \phantom{0}5391.3 & 0.082 & 13.7   \\
    \texttt{GBRT} & 10280.1 & 0.085 & 22.5  \\
    \texttt{RF} &  10657.5 & 0.114 & 32.6  \\
    \texttt{Adaboost} & 10021.2 & 0.075 & 20.9 \\
    \texttt{DNF} & \phantom{0}2326.2 & 0.053 & \phantom{0}9.3  \\
    \texttt{frankenz} & \phantom{0}7807.7 & 0.069 & 22.0  \\
    \texttt{NNPZ} & 10112.7 & 0.082 & 22.4  \\
    \hline
    \end{tabular}
    \label{tab:statsWeight}
\end{table}

Another sample being somewhat representative of the full photometric one is the shear sample with the 30-band photo-$z$s from \cite{Laigle2016}. Using these redshifts as reference, the scatter $\sigma_{\rm{30bands}}$ and outlier fraction $\eta_{\rm{30bands}}$ are presented in Table~\ref{tab:stats30bands}. The number of selected sources $N_{\rm{30bands}}$ for each method is also shown.

\begin{table}[h]
    \centering
    \caption{Point estimate statistics for the L15 sample. The number of sources $N_{\rm{30bands}}$, scatter $\sigma_{\rm{30bands}}$, and outlier fractions $\eta_{\rm{30bands}}$ are provided.}
    \begin{tabular}{lcccc}
    \hline\hline
     & $N_{\rm{30bands}}$ & $\sigma_{\rm{30bands}}$ & $\eta_{\rm{30bands}}$   \\
    &  & & [\%]&  \\
    \hline
    \rule{0pt}{1.2em}\texttt{Le Phare} & 36842 & 0.055 & 11.7 \\
    \texttt{CPz}  & 43258 & 0.08\phantom{0} & 20.0   \\
    \texttt{Phosphoros}  & 38649 & 0.069 & 16.5  \\
    \texttt{EAzY} & 25114 & 0.059 & \phantom{0}9.8  \\
    \texttt{METAPHOR} & 13830 & 0.04\phantom{0} & \phantom{0}2.7  \\
    \texttt{ANNz}& 52094 & 0.114 & 32.3 \\
    \texttt{GPz} & 23207 & 0.082 & 13.7  \\
    \texttt{GBRT} & 50980 & 0.081 & 19.1  \\
    \texttt{RF} & 52391 & 0.136 & 37.4  \\
    \texttt{Adaboost} & 49718 & 0.096 & 26.9  \\
    \texttt{DNF} & \phantom{0}9694 & 0.052 & \phantom{0}8.5  \\
    \texttt{frankenz} & 42808 & 0.07\phantom{0} & 19.1  \\
    \texttt{NNPZ} & 51047 & 0.081 & 19.9  \\
    \hline
    \end{tabular}
    \label{tab:stats30bands}
\end{table}

\begin{figure*}
    \centering
    \includegraphics[width=\linewidth]{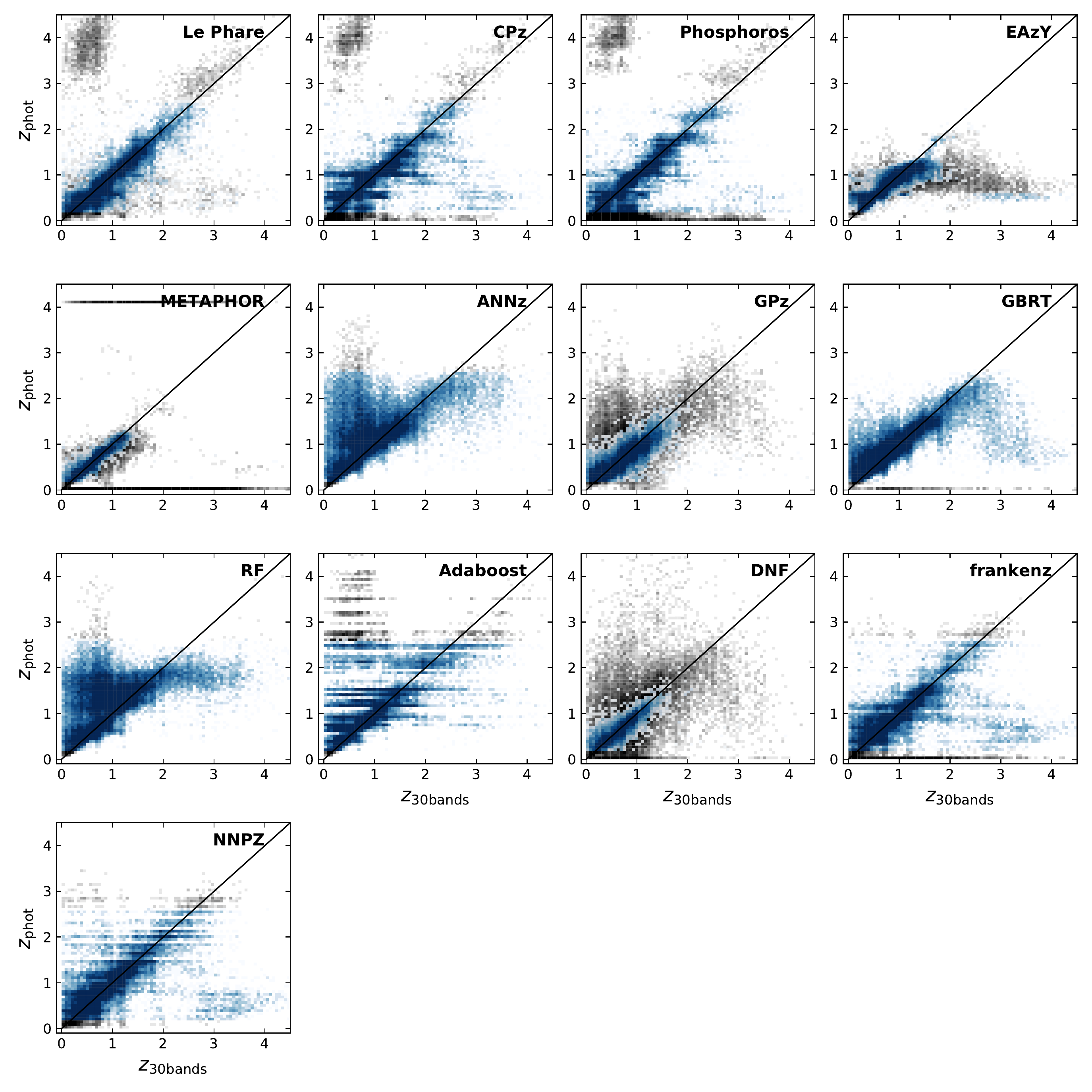} \\
    \includegraphics[width=\linewidth]{colorbarDensitySpecz02-26.pdf}
    \caption{Photometric redshifts between $0.2<z\leq2.6$ measured with all the methods compared to the \citet{Laigle2016} 30-band photometric redshifts. The color code is the same here as in Fig.~\ref{fig:densityPlot}, meaning that the shades of blue represent the \textit{Euclid} selection, and the shades of gray represent the rest of the L15 sample. As in Fig.~\ref{fig:densityPlot}, undefined or negative point estimate values have been set to 0 in the plots.}
    \label{fig:densityPlot30bands}
\end{figure*}

\section{photo-$z$ versus 30-band photo-$z$}
\label{sec:30bandsPointEstimates}

photo-$z$s provided by all the tested methods are compared to reference redshifts to examine the performance of the codes. In Sect~\ref{sec:results-pointEstimates}, we present a comparison between spec-$z$s and photo-$z$s, specifically shown in Fig.~\ref{fig:densityPlot}. Figure~\ref{fig:densityPlot30bands} makes a comparison between the code photo-$z$s and 30-band photo-$z$s of \cite{Laigle2016}, which better represent the full photometric sample. The resulting metrics associated with these plots are presented in Table~\ref{tab:stats30bands}.

\section{CRPS plots}

\begin{figure*}
    \centering
    \includegraphics[width=\linewidth]{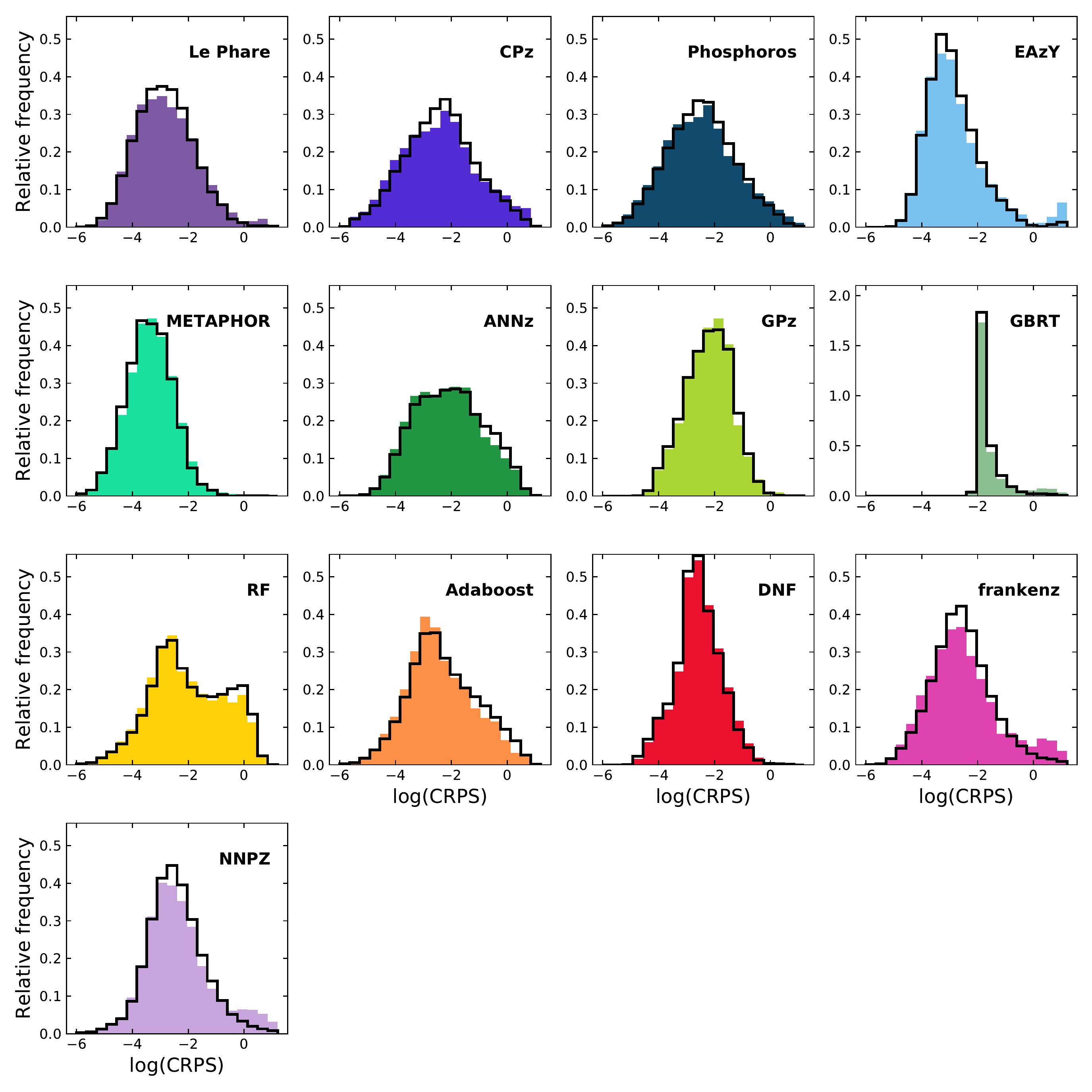}
    \caption{CRPS plots for all the methods. Colored histograms are the histograms of log(CRPS) for the weighted  spectroscopic sample, while solid black lines are the histograms for the L15 sample.}
    \label{fig:crps}
\end{figure*}

In Sect~\ref{sec:results-PDFs}, we present the mean and the median continuous ranked probability score (CRPS) for all the methods, specifically in Table~\ref{tab:CRPS}. In Fig.~\ref{fig:crps} we show the full distributions of CRPSs, for both the spectroscopic and the L15 samples. The spectroscopic sample CRPSs have their distribution weighted by the color-space weights (see Sect~\ref{sec:results-pointEstimates}).

\section{Fraction of good sources per bin}

\begin{figure}
    \centering
    \includegraphics[width=\linewidth]{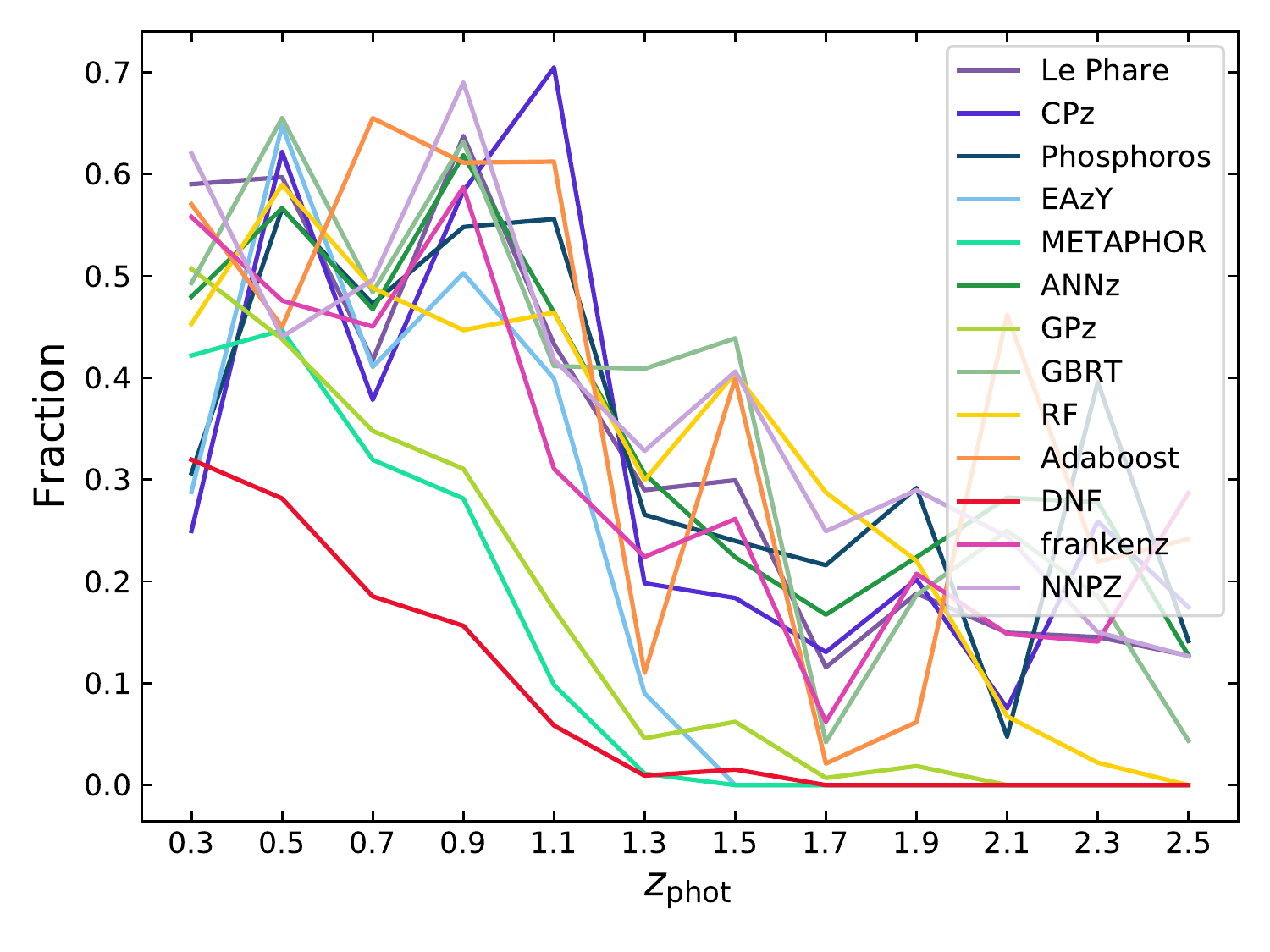}\\
    \includegraphics[width=\linewidth]{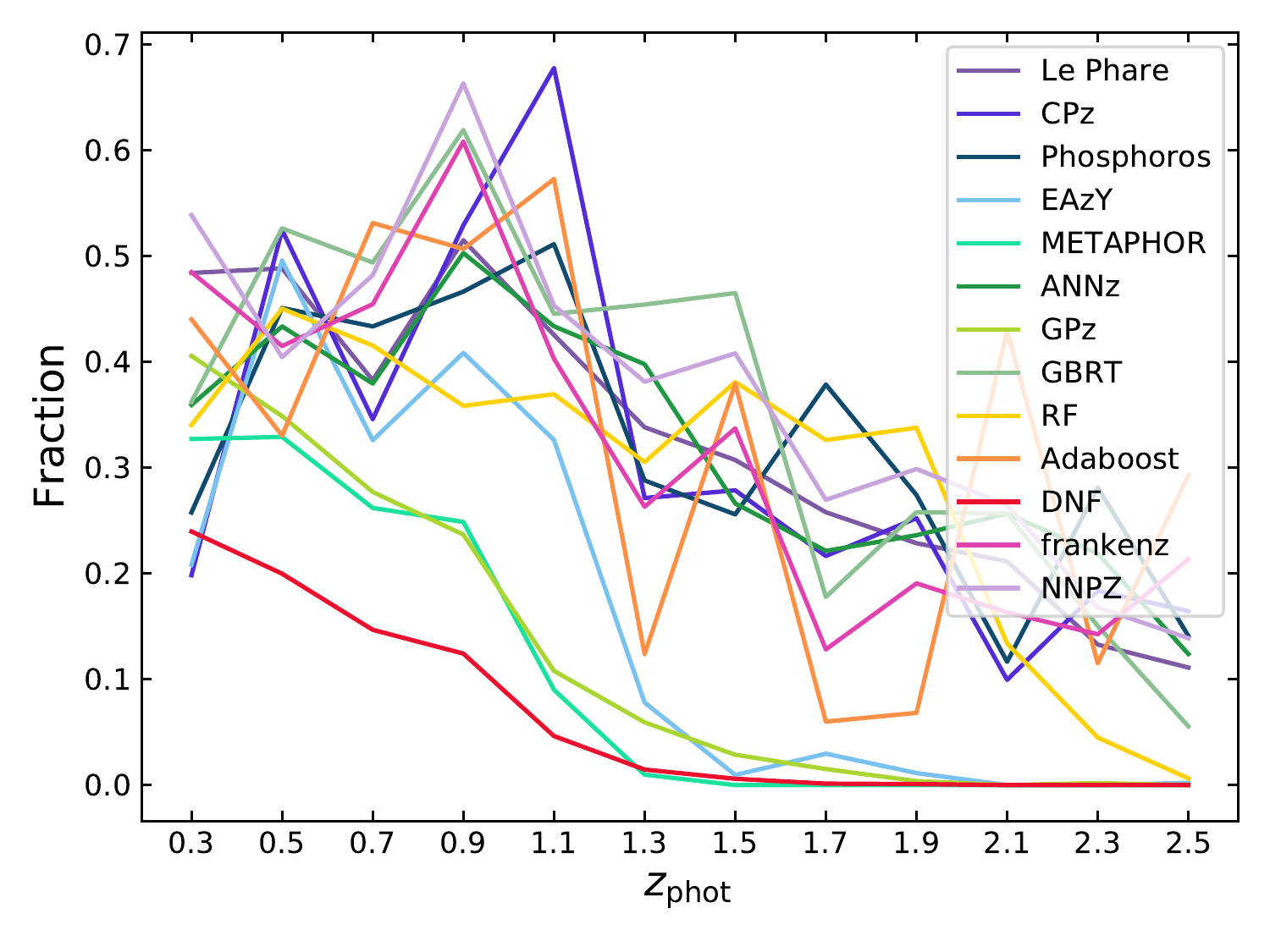}
    \caption{Fraction of sources per redshift bin that have both photo-$z$ and true $z$ belonging to the bin, compared to the number of sources with their spec-$z$s in the bin. \textit{Top}: true $z$ from GD{the} weighted spectroscopic sample. \textit{Bottom}: true $z$ from L15. }
    \label{fig:goodfraction}
\end{figure}

In Sect.~\ref{sec:discussion-metrics} we discuss which metrics we should consider to maximize the figure of merit of the weak-lensing signal. In Eq.~(\ref{eq:score3}), we correct the $F_{0XX}$ metrics in all the bins by the square root of the fraction of sources appropriately attributed to the considered bin. Figure~\ref{fig:goodfraction} show this fraction in all the bins, for all the methods, using both the weighted spectroscopic sample (top panel) and the L15 sample (bottom panel). These values were used to compute the metrics presented in Fig.~\ref{fig:metric}.

\end{appendix}

\end{document}